\documentclass[journal]{IEEEtran}
\usepackage{amssymb,epsfig,color,cite,amsmath,subfigure,amsfonts,fancybox,mathrsfs}
\usepackage{balance}

\usepackage{algorithm}
\usepackage{algpseudocode}

\usepackage{array}
\newcolumntype{P}[1]{>{\centering\arraybackslash}p{#1}}

\usepackage{float}

\newcommand{\wdt}{\widetilde}
\newcommand{\ol}{\overline}
\newcommand{\wdh}{\widehat}

\newcommand{\bed}{\begin{displaymath}}
\newcommand{\eed}{\end{displaymath}}
\newcommand{\bea}{\bed\begin{array}{rl}}
\newcommand{\eea}{\end{array}\eed}
\newcommand{\beq}[1]{\begin{equation} \label{#1}}
\newcommand{\eeq}{\end{equation}}
\newcommand{\barray}{\begin{array}{ll}}
\newcommand{\earray}{\end{array}}
\newcommand{\disp}{\displaystyle}
\newcommand{\ad}{&\!\!\!\disp}

\newcommand{\e}{{\epsilon}}

\def\clC{{\cal C}}
\def\clD{{\cal D}}
\def\clE{{\cal E}}

\def\clN{{\cal N}}

\def\clR{{\cal R}}
\def\clS{{\cal S}}

\def\clU{{\cal U}}
\def\clV{{\cal V}}

\def\e{{\epsilon}}

\def\Base{{\rm Base}}

\def\Range{{\rm Range}}

\def\Rank{{\rm Rank}}

\newcommand{\dbU}{{\Bbb U}}

\def\one{{\hbox{1{\kern -0.35em}1}}}

\newtheorem{thm}{Theorem}

\newtheorem{lem}{Lemma}
\newtheorem{rem}{Remark}

\newtheorem{exm}{Example}
\newtheorem{asm}{Assumption}

\begin{document}
\title{Contingency Detection in Modern Power Systems: A Stochastic Hybrid System Method}

\author{Shuo Yuan,\thanks{Shuo Yuan is with the Department of Electrical and Computer Engineering,
	Wayne State University, Detroit, Michigan 48202, USA {\tt\small <shuoyuan@wayne.edu>}}
Le Yi Wang,\thanks{Le Yi Wang is with the Department of Electrical and Computer Engineering,
	Wayne State University, Detroit, Michigan 48202, USA {\tt\small <lywang@wayne.edu>}} \IEEEmembership{Life Fellow, IEEE,}
George Yin,\thanks{George Yin is with the Department of Mathematics,
University of Connecticut,
Storrs, Connecticut 06269-1009, USA {\tt\small
<gyin@uconn.edu>}} \IEEEmembership{Life Fellow, IEEE,}
Masoud H. Nazari,\thanks{Masoud Nazari is with the Department of Electrical and Computer Engineering,
Wayne State University, Detroit, Michigan 48202, USA {\tt\small <masoud.nazari@wayne.edu>}}
\IEEEmembership{Senior Member, IEEE}

\thanks{}}
\date{}
\maketitle

\begin{abstract}
This paper introduces a new stochastic hybrid system (SHS) framework for contingency detection  in modern power systems (MPS). The framework uses stochastic hybrid system representations in state space models to expand and facilitate capability of contingency detection. In typical microgrids (MGs), buses may contain various synchronous generators, renewable generators, controllable loads, battery systems, regular loads, etc. For development of SHS models in power systems, this paper introduces the concept of dynamic and non-dynamic buses. By converting a physical power grid into a  virtual linearized state space model and representing contingencies as random switching of system structures and parameters, this paper formulates the contingency detection problem as a joint estimation problem of discrete event and continuous states in stochastic hybrid systems. This method offers unique advantages, including using common measurement signals on voltage and current synchrophasors to detect different types and locations of contingencies, avoiding expensive local direct fault measurements and detecting certain contingencies that cannot be directly measured. The method employs a small and suitably-designed probing signal to sustain the ability of persistent contingency detection. Joint estimation algorithms are presented with their proven convergence and reliability properties. Examples that use an IEEE 5-bus system demonstrate the main ideas and derivation steps.  Simulation case studies on  an IEEE 33-bus system are used for detecting transmission line faults and sensor interruptions.
\end{abstract}


\section{Introduction}

Reliability of modern power systems (MPS) mandates fast and accurate detection of cyber-physical contingencies in diversified categories, including line faults, generator failures, sensor malfunctions, communication system disruptions, among many others \cite{KN}. 
Contingency detection in power systems is a critical and challenging task.  Power system contingencies include physical faults such as transmission line faults, information failures such as communication channel interruptions and sensor faults, and intentional cyber attacks that alter system structures and parameters.

Traditionally, unique and dedicated local sensors and switching circuits have been designed and implemented at numerous locations on transmission lines, buses, generators, and users. These detection devices are exemplified by power line sensors, faulty circuit indicators, fault passage indicators, over-current relays, and so on. Enhancement of reliability can also be achieved by using existing sensor systems with advanced detection methodologies and algorithms \cite{Sin2011Trans, 
Nan2018Fault, Fu2022Ele}.

This paper introduces a new approach that employs system dynamics and their switching for enhancing and expanding the capability of contingency detection. This approach offers several appealing advantages. For example, an existing sensor such a phasor measurement unit (PMU) or a current sensor on a bus can be used to detect many types and different locations of contingencies. This much enhanced capability is achieved by embedding contingencies within dynamic systems so that a switching from a normal operating condition to a faulty condition caused by a contingency can be detected as two different dynamic systems. This leads to the stochastic hybrid system (SHS) framework of power systems. This new framework for modeling power systems was introduced in the companion paper \cite{cd1} that treats also state estimation problems. This paper is focused on contingency detection. Since this approach must rely on system dynamics to detect  contingency and estimate state simultaneously, they face numerous challenges. This paper aims to resolve these challenges  and introduce useful algorithms in the SHS framework.

This paper considers power grids whose buses may contain various traditional synchronous generators and renewable generators, controllable loads, energy storage systems, battery systems, regular loads, and so on. Their system dynamics are represented by nonlinear state space models. Contingencies are represented by jumps in system structures (such as loss of a transmission line, loss of communication channel in an interval), model parameters (such as impedance jumps in transmission lines and generation parameters), loss of a member (loss of a load cluster on a bus), among many other scenarios. Most power system contingencies occur randomly. Adding such random jumps into dynamic models of MGs, the system models become stochastic state-space hybrid systems.

Due to their critical importance,  contingency detection and state estimation problems in power systems have been investigated extensively. For example, the reference \cite{Gop2015Trans} proposed a support vector machine-based fault localization methodology to identify and localize transmission line faults occurring at any location in a power grid based on PMUs (phasor measurement units) measurements.  
A method of fault detection and classification was presented in \cite{Chen2018Detection} for power transmission lines based on convolutional sparse autoencoder. A robust fault detection and discrimination technique for transmission lines was proposed in \cite{Aff2018A}, which utilizes a robust method of phasor estimation to compute accurate fault impedance along with a feature value extracted from the samples of voltage and current signals. A new algorithm was  introduced in \cite{Gho2020Detecting}   for short-circuit fault detection and identification based on state estimation taking into account the measurements in active distribution systems. 

In contrast to the aforementioned references, this paper provides a different method by using
SHS.
Within the SHS framework, this paper is focused on exploring the potential of using common sensors such as PMUs or frequency or voltage  for detecting contingency and estimating internal states jointly, for enhancing reliability and expanding the capability of contingency detection. To the best of our knowledge, this paper is the first effort in applying the SHS to detect contingencies in power systems. Our approach is based on the theoretical foundation of our recent papers \cite{WY1,WY2}. It should be emphasized that as a newly developed theoretical work, applications of the results from \cite{WY1,WY2} in power systems are highly challenging, including derivation of virtual dynamic SHS models, their linearization, algorithm implementation, 
and convergence validation, etc.   By using collaboratively dynamic hybrid system models, stochastic information on system jumps, and advanced observer design methods, this paper achieves contingency detection and state estimation simultaneously.

The main contributions of this paper are summarized as follows:
\begin{enumerate}
 

\item It introduces an approach of using SHS for joint contingency detection and state estimation. The interwinding nature of continuous state and switching processes makes it necessary to perform joint estimation for contingency detection. By employing the rich information from the dynamic system models, it becomes possible to jointly detect contingencies and estimate the continuous states by using only limited numbers of sensors. The joint estimation problem is much more complicated than the state estimation problems in power systems since they assume the dynamic system is known.
	
\item It introduces a design method for selecting an input probing such that contingencies of different types and locations can be detected by using only a limited set of sensors. A mode-modulated input design method for suitably selecting inputs is presented so that detectability on contingencies  can be persistently sustained. 

\item It develops a two-time-scale framework and algorithms for jointly detecting contingencies and estimating the continuous states simultaneously. Convergence properties of the algorithms are established.
	
\item It employs two common IEEE testing systems to validate and evaluate  models, detection algorithms, observer design, convergence properties, 
and algorithm robustness. The methods of this paper are highly scalable. The complexity of the virtual dynamic SHS depends on the number of dynamic buses that can be numerically derived using commercial software packages of power flow analysis such as MATPOWER. These numerical methods have been used in case studies on the IEEE 33-bus system.
\end{enumerate}

The paper is organized as follows. Section \ref{P} defines notations and the main problems of this paper. Section \ref{SSM} derives state space models of MGs. Sensor systems, contingencies, and stochastic hybrid systems are described in Section \ref{sensor}. Section \ref{Design} presents observer design procedures and detection algorithms, and establishes convergence. Performance evaluation case studies are discussed in Section \ref{case}. 
 The main conclusions of this paper are summarized in Section \ref{Conc}.

\section{Preliminaries}\label{P}

For a column vector $v\in \mathbb{R}^n$, $\|v\|$ is its Euclidean norm. For a matrix $M\in \mathbb{R}^{n\times m}$, $M^{\top}$ is its transpose, $\lambda(M)$ is an eigenvalue of $M$,  and $\sigma(M)=\sqrt{\lambda(M^{\top}M)}$ is a singular value of $M$.
The  kernel or null space of $M\in \mathbb{R}^{n\times m}$ is
$\ker(M)=\{x\in \mathbb{R}^m: Mx=0\}$ and its range is $\Range(M)=\{y=Mx: x\in \mathbb{R}^m\}$.
For a subspace $\dbU\subseteq \mathbb{R}^n$ of dimension $p$, a matrix $M\in \mathbb{R}^{n\times p}$ is said to be a base matrix of $\dbU$, written as $M=\Base(\dbU)$,  if the column vectors of $M$ are linearly independent and ${\rm Range}(M)=\dbU$. A function $y(t)\in \mathbb{R}$ in a time interval $[0,\tau)$ is piecewise continuously differentiable if $[0,\tau)$ can be divided into a finite number of subintervals $[t_{k-1},t_k)$, $k=1,\ldots, \ell$, $t_0=0$, $t_\ell =\tau$ such that $y(t)$ is right continuous in  $[t_{k-1}, t_k)$ and continuously differentiable, to any order as needed, in  $(t_{k-1}, t_k)$. The space of such functions is denoted by $\clC[0,\tau)$.

For an AC power microgrid, all voltages and currents  will be represented by their phasors $\vec V=V \angle \delta$ and $\vec I=I \angle \gamma$. Sensors in power systems are highly diversified, including
 PMUs, frequency, voltage, power  measurements, signal transducers for protection, rotational speed, torque, temperature,  among many others. Furthermore, communication systems are used for data transmission.  The microgrid  can be viewed as a networked system with $\gamma$ buses connected by transmission lines. This network system is represented by a graph $\clN=\{\clV, \clE\}$ where $\clV$ is the set of buses (vertices in a graph) and $\clE$ is the set of feeder/transmission links (edges in a graph). The transmission line $(i,j)\in \clE$ is bi-directional, i.e., $(i,j)\in \clE \rightarrow (j,i)\in \clE$.
For Bus $i$, its neighbor $\clN_i$ is the set of buses $j$ that are connected to it, namely, $\clN_i = \{j\in \clV: (i,j)\in \clE \hbox{ or } (j,i)\in \clE \}$. By default, $(i,i)\in \clE$.

\section{State Space Models of Microgrids}\label{SSM}

We now summarize the main SHS framework introduced and detailed in \cite{cd1}. We should emphasize that the framework is highly general. For demonstration, we will use real power management  problems in case studies.
Power systems are highly complicated and interconnected systems. Microgrids are unique in which power generations can step from traditional synchronous generators, power-electronics-based wind turns, solar panels, battery systems, and controllable loads.
To derive a state space model representation of SHS of power systems, it is essential to characterize bus types according to their dynamics. Consequently, we divide buses into two types: dynamic buses and non-dynamic buses. This classification is independent of traditional classifications such as PV/PQ buses or dispatchable/non-dispatchable buses.

\subsection{Dynamic Buses}

If Bus $i$ is dynamic, then it is represented by a local state space model,\vspace{-3pt}
\begin{small}
    \beq{dy00}  \dot z^d_i = f_i(z^d_i, z_i^-, v_i^d, \ell^d_i),\eeq 
\end{small}
where $z^d_i$ is the local state variable, $z_i^-$ is the neighboring variables of Bus $i$ which may be state variables of its neighboring dynamic buses, or intermediate variables of its neighboring non-dynamic buses, $v_i^d$ is the local  control input, and $\ell^d_i$ is the local congregated total load that cannot  be actively controlled, such as regular loads, fixed-blade wind generators, solar panels, constant-charging-current batteries, etc. The control input is set of controllable (i.e., dispatchable) variables such as  generator mechanical power input, controllable loads, actively managed battery systems, tunable wind turbines, etc.  If a bus does not have any dispatchable assets, then $v_i^d=0$.

\subsection{Non-dynamic Buses}

If the $j$th bus is non-dynamic, which is in a steady state or pseudo-steady state,  then it is represented by an implicit algebraic relationship,
\vspace{-3pt}\beq{nd0} 0=g_j(z^{nd}_j,z_j^-, v_j^{nd},\ell^{nd}_j),\vspace{-2pt}\eeq
where $z^{nd}_j$ is the local state variable vector, $z_j^-$ is the neighboring variables, $v_j^{nd}$ is the local control input, and $\ell^{nd}_j$ is the local load.

\begin{rem}

\begin{enumerate}
 
\item  Dynamic/non-dynamic designation is related to the local system's models. They do not affect the designation of power flow analysis such as PV, PQ, slack buses. For example, a load is typically considered as a PQ bus. If the load is motor with its own dynamic model, then the bus is a dynamic bus.
\item  Traditionally, generators are dynamic. But fast reaction power sources like batteries, may be simplified as non-dynamic and represented by their near-steady-state algebriac relationships. For the same token, loads can be either static (non-dynamic) or dynamic. The ZIP and exponential load models are static. But induction motors are usually modeled dynamic systems, and so are exponential recovery load (ERL) models.
\item  Dynamic/non-dynamic bus classification can change when local sensors and controllers are included. For example, PID controllers are dynamic systems and after applying them to control a non-dynamic system on a bus, the bus becomes dynamic.
\item  Dynamic/non-dynamic designation also does not affect  whether a bus is dispatchable or not. A dispatchable resource has controllable real or reactive  powers to participate in control or market of a power grid. It is represented as part of the control input $v_i^d$ in (\ref{dy00}).
\end{enumerate}
\end{rem}

The general nonlinear state equation (\ref{dy00}) is highly versatile in representing dynamic systems on a bus. For example, this may be a common swing equation for synchronous generators\vspace{-3pt}
    \beq{dy0}
M_i \dot \omega_i +g_i( \omega_i) = P_i^{in}-P_i^L-P_i^{out},
\eeq\vspace{-1pt}
where  $\delta_i$ is its electric angle, $\omega_i=\dot \delta_i$, $M_i$ is the equivalent electric-side inertia, $g_i(\cdot)$ represents the nonlinear damping effect, and $g_i(\cdot)$ is continuously differentiable 
satisfying
$\omega_i g_i(\omega_i)> 0$ for $\omega_i \neq 0$. Linearization of $g_i(\cdot)$ around  $\omega_i=0$ is
$b_i \omega_i$ with  $b_i > 0$. Also, $P_i^{out}$ is the total transmitted power from Bus $i$ to its neighboring buses.
As an extension, if it is required to include steam turbine control with generators, the dynamic model will combine both turbine and generator dynamics and state variables will then include flow rate and other mechanical system variables.
However, if a bus has a renewable generator or a battery system with power electronic based control mechanisms, then their dynamic models will be different, inherited from the specific dynamic models derived for such physical systems.

For both dynamic and non-dynamic buses, the interaction of the local variables with their neighboring buses is based on the standard power flow relationships.
    Suppose that the transmission line between Bus $i$ and Bus $j$ has impedance $X_{ij}\angle \theta_{ij}$.   The line current is\vspace{-3pt}
 \begin{small}
   \begin{align*}
I_{ij} \angle \gamma = {V_i\angle \delta_i \!-\! V_j\angle \delta_j \over X_{ij}\angle \theta_{ij}}
 = {V_i\over X_{ij}} \angle (\delta_i \!-\!\theta_{ij}) - {V_j\over X_{ij}} \angle (\delta_j\!-\!\theta_{ij}).
 \end{align*}
 \end{small}
Denote $\delta_{ij}=\delta_i-\delta_j$.
The complex power flow from Bus $i$ to Bus $j$ at Bus $i$ is
\begin{small}
   \begin{align*}S_{ij} = V_i \angle \delta_i \times I_{ij}\angle (-\gamma) =
{V_i^2\over X_{ij}} \angle \theta_{ij} - {V_iV_j\over X_{ij}} \angle (\theta_{ij}+\delta_{ij}), \end{align*}
 \end{small} 
which implies that the transmitted real and reactive powers at Bus $i$ are 
$$
P_{ij} \!=\! {V_i^2\over X_{ij}} \cos( \theta_{ij}) \!-\! {V_iV_j\over X_{ij}}\cos (\theta_{ij}\!+\!\delta_{ij}),$$
$$
Q_{ij}\! =\! {V_i^2\over X_{ij}} \sin( \theta_{ij}) \!-\! {V_iV_j\over X_{ij}}\sin (\theta_{ij}\!+\!\delta_{ij}).
$$


\subsection{Virtual Dynamic State Space Models}

Suppose that the $\gamma$ buses in $\clN$ contain $\gamma^d$ dynamic buses and $\gamma^{nd}=\gamma-\gamma^d$ non-dynamic buses.\footnote{Since this paper deals with state estimation under state space models, we assume that $1\leq \gamma^d\leq \gamma$, namely at least one bus is dynamics. But $\gamma^{nd}=0$ is possible, meaning that all buses are dynamic.} Without loss of generality, let the first $\gamma^d$ buses be dynamic.
Define the states, inputs, loads of dynamic buses and non-dynamic buses as \vspace{-2pt}
$$
z^{\!d}\!\!=\!\!\! \begin{bmatrix}\!\!
     z^d_1\!\!\\ \!\!\vdots \!\!\\ \!z^d_{\gamma^d} \!\!\end{bmatrix}\!\!,
 v^{\!d}\!\!=\!\!  \begin{bmatrix}\!\!v_1^d \!\!\\ \!\! \vdots \!\!\\ \!v_{\gamma^d}^d \!\!\end{bmatrix}\!\!,
   \ell^d \!\!=\!\! \begin{bmatrix} \!\!\ell^d_1\!\! \\ \!\!\vdots \!\!\\ \!\ell^d_{\gamma^d}  \!\!\end{bmatrix} $$ and $$ z^{\!nd} \!\!=\!\!  \begin{bmatrix} \!z^{nd}_{\!\gamma^d\!+\!1}\!\!\! \\ \vdots \\ z^{nd}_{\gamma} 
\end{bmatrix}\!\!,
 v^{\!nd}\!\! =\!\!  \begin{bmatrix}\!v_{\!\gamma^d\!+\!1}^{nd}\!\!  \\ \vdots \\ v_{\gamma}^{nd} \end{bmatrix}\!\!,
\ell^{nd}\!\!=\!\!  \begin{bmatrix} \!\ell^{nd}_{\!\!\gamma^d\!+\!1} \!\!\\ \vdots \\ \ell^{nd}_\gamma \end{bmatrix}\!\!.
$$
Define the state, input, and load of all buses as
 $
z \!=\!  \begin{bmatrix} \!\!z^d \!\!\\  z^{nd} \end{bmatrix},
v =  \begin{bmatrix} v^{d} \\  v^{nd} \end{bmatrix},
\ell= \begin{bmatrix} \ell^d \\  \ell^{nd}\end{bmatrix}.$ 
By (\ref{nd0}),
for non-dynamic buses, we have 
\[
G^0(z^{nd}, z^d, v^{nd},\ell^{nd}) 
\!=\! \begin{bmatrix}g_{\gamma^d+1}(z^{nd}_{\gamma^d+1},z_{\gamma^d+1}^-, v_{\gamma^d+1}^{nd},\ell^{nd}_{\gamma^d+1})\\
\vspace{-2pt}\vdots\vspace{-2pt}\\
g_{\gamma}(z^{nd}_{\gamma},z_{\gamma}^-, v_{\gamma}^{nd}, \ell^{nd}_{\gamma})
\end{bmatrix}\!\!=\!0. 
\]
For physical power grids, given $z^d$, $v^{nd}$, $\ell^{nd}$, this equation has a unique solution within permitted operating ranges, leading to the symbolic relationship$
z^{nd}= H(z^d,v^{nd},\ell^{nd}).
$
Furthermore, by the dynamic systems in (\ref{nd0}),
$
\dot z^d = F^0(z^d, z^{nd},v^d,\ell^d).
$
Then, we obtain
\begin{small}
    \begin{align}\label{total}
\!\!\!\dot z^d \!= F^0(z^d,H(z^d,v^{nd},\ell^{nd}),v^d,\ell^d)=F(z^d,v^{d},v^{nd},\ell^{d},\ell^{nd}).
\end{align}
\end{small}

\subsection{Linearization}

In power system control problems, it is common to linearize the nonlinear dynamics (\ref{total})  near nominal operating points \cite{KN,GS}. The linearization process involves the following standard steps. Given the steady-state loads $\overline \ell=[\overline \ell^{d}, \overline \ell^{nd}]^{\top}$ and steady-state input real powers $\overline v=[\overline v^{d}, \overline v^{nd}]^{\top}$, the steady-state
$\overline z^d$ (equilibrium point or the nominal operating condition) is the solution to
$F(\ol z^d,\ol v^d,\ol v^{nd},\overline \ell^{d}, \overline \ell^{nd}) = 0.$

By defining the perturbation variables from their nominal values as
$x=z^{d}-\overline z^{d}, u=v^d-\overline v^d, u^{n}=v^{nd}-\overline v^{nd}, \zeta=\ell^{d}-\overline \ell^{d}, \zeta^n=\ell^{nd}-\overline \ell^{nd}$,
the linearized system is\vspace{-3pt}
\beq{total2}
\dot x = A x+B_1 u+ B_2 u^n + D_1 \zeta+ D_2 \zeta^n,
\eeq
where the matrices are  the related Jacobian matrices\begin{small}
$
A = {\partial F(\cdot)\over \partial z^d},
B_1 = { \partial F(\cdot)\over \partial v^{d} }, 
B_2=  {\partial F(\cdot)\over \partial v^{nd}} ,  
D_1=  {\partial F(\cdot)\over \partial \ell^{d} }, 
D_2 =  {\partial F(\cdot)\over \partial \ell^{nd} }
$ 
\end{small} at $z^d=\ol z^d, v=\ol v, \ell=\ol \ell.$


\begin{exm}\label{exm1} {\rm
Suppose that the generators can be represented by the swing equations\vspace{-3pt}
\bea
M_i \dot \omega_i +g_i( \omega_i) = P_i^{in}-P_i^L-P_i^{out},\vspace{-5pt}
\eea
where  $\delta_i$ is its electric angle, $\omega_i=\dot \delta_i$, $M_i$ is the equivalent electric-side inertia, $g_i(\cdot)$ represents the nonlinear damping effect, and $g_i(\cdot)$ is continuously differentiable 
satisfying
$\omega_i g_i(\omega_i)> 0$ for $\omega_i \neq 0$. Linearization of $g_i(\cdot)$ around  $\omega_i=0$ is
$b_i \omega_i$ with  $b_i > 0$. Also, $P_i^{out}$ is the total transmitted power from Bus $i$ to its neighboring buses, i.e.,
\begin{small}
    $
P_i^{out} = \sum_{j\in \clN_i} P_{ij} =\sum_{j\in \clN_i} \left[{V_i^2\over X_{ij}} \cos( \theta_{ij}) - {V_iV_j\over X_{ij}}\cos (\theta_{ij}+\delta_{ij})\right] =\sum_{j\in \clN_i} q(\delta_i,\delta_j),
$
\end{small}
where
\begin{small}$
q(\delta_i,\delta_j)={V_i^2\over X_{ij}} \cos( \theta_{ij}) - {V_iV_j\over X_{ij}}\cos (\theta_{ij}+\delta_{ij}).
$
\end{small}
The two-bus system shown in Fig. \ref{connection} has $\theta_{12}=90^o$ which is the angle of impedance, namely the transmission line is lossless. Both buses are dynamic dispatchable  buses  with state variables
$
z_1^d=[\delta_1; \omega_1], z_2^d=[\delta_2; \omega_2].
$
Suppose that $g_1(\omega_1)=b_1 \omega_1$, $b_1>0$, and $g_2(\omega_2)=b_2 \omega_2$, $b_2>0$. Denote $\beta=\beta_{12}= V_1 V_2/X_{12}$ and $\delta = \delta_1-\delta_2$. Then,
\begin{small}
$
f_1(z_1^d,z_2^d) = \left[\omega_1; - {b_1 \omega_1 \over M_1} -{1\over M_1}  \beta\sin (\delta)\right],
f_2(z_2^d,z_1^d) = \left[\omega_2; - {b_2 \omega_2 \over M_2} -{1\over M_2}  \beta\sin (-\delta)\right].
$    
\end{small}
\vspace{-11pt}
\begin{figure}[htbp] \vspace{-10pt}
	\setlength{\unitlength}{0.1 in}
	\begin{center}
		\includegraphics[height=4cm]{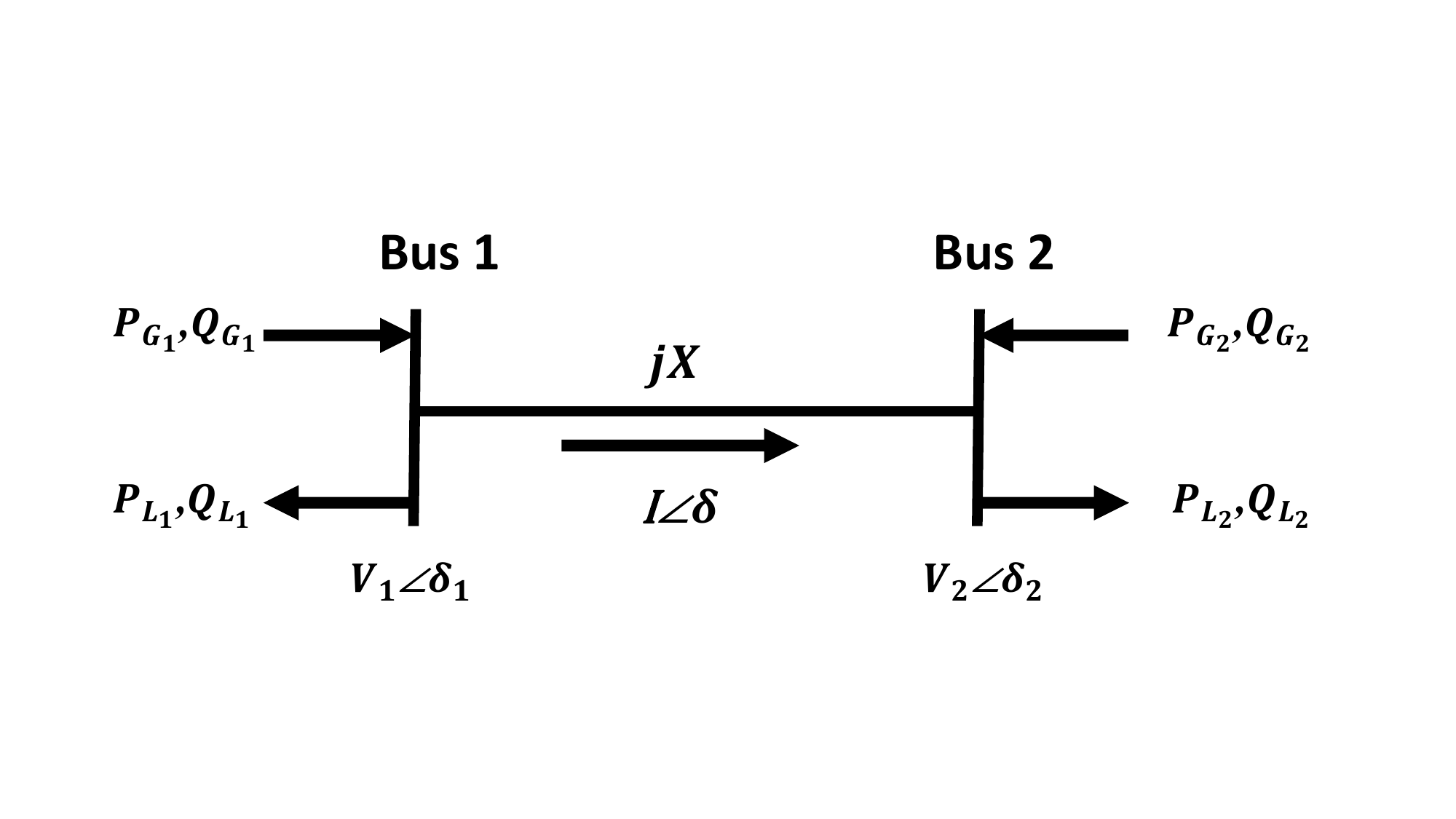} \vspace{-8pt}
	    \caption{A link in microgrids}
 \label{connection} 
	\end{center}
\end{figure}

Given $\overline v_1=P_1^{in}$, $\overline v_2=P_2^{in}$, $\overline \ell_1^d=P_1^{L}$, $\overline \ell_2^d=P_2^{L}$, the equilibrium point is
is $\overline{\omega}_1=0$, $\overline{\omega}_2=0$, and
\begin{small}
$
\overline \delta = \sin^{-1}\left({P_1^{in}-P_1^{L}\over \beta}\right).
$    
\end{small}
Assume $M_1=1$, $M_2=1.5$, $b_1=0.2$, $b_2=0.31$, $\beta=200$, $P_1^{in}=100$, $P_2^{in}=50$, $P_1^{L}=70$, and $P^2_L=80$. Then, the equilibrium point is $\overline\delta = 0.1506$ (rad). Under these given values, the linearized system is (\ref{total2}) with\vspace{-3pt}
  \beq{Amatrix}
A =\begin{bmatrix}
0   & 1      &   0    &     0\\
-197.7372  & -0.2 & 197.7372     &    0\\
0   &      0    &     0  &  1\\
131.8248    &     0  & -131.8248  & -0.2067
\end{bmatrix},
\eeq     
$B_1\! =\begin{bmatrix}
    0 \!\!&\!\! 0\\
1/M_1 \!\!&\!\! 0\\ 0 \!\!&\!\! 0 \\ 0 \!\!&\!\!1/M_2
\end{bmatrix} \! ,D_1 \!=\begin{bmatrix}  0 \!\!\!&\!\!\! 0\\
-1/M_1 \!\!\!&\!\!\! 0\\ 0 \!\!\!&\!\!\! 0 \\ 0 \!\!\!&\!\!\! -1/M_2
\end{bmatrix} \! , B_2\!=0, D_2\!=0$.
 }
\end{exm}

 \section{Contingencies  and Stochastic Hybrid System Models}\label{sensor}

\subsection{Sensor Systems and Observability}
\vspace{-1pt}
For power system operation and contingency detection, many sensors must be deployed, such as voltages, frequencies, PMUs, over-current protection transducers, among many others. Sensor selection and placement are important for managing SHS. Mathematically, sensor choice (which variable to measure) and location (which bus to measure) are reflected in the output equation (\ref{total2}).

Adding a sensor system with $y=Cx$ where $C$ is the sensing matrix, we have the following state space model: 
     \beq{total3}
\left\{\begin{array}{rcl}
\dot x &=& A x+B_1 u+B_{2}u^n+ D_1 \zeta + D_{2}\zeta^{n},\\
y& = & Cx.
\end{array}
\right. 
\eeq

\begin{rem}
 The system (\ref{total3}) is a linearized system whose variables are perturbations from their nominal values. Physically, $y$ is the difference between the measured value $y_{\text{measured}}$ and the nominal value $y_{\text{nominal}}$ at the operating point.
  For example, the phasors on buses can be measured by PMUs. Suppose that only $\delta_1$ is measured, this can be represented by $y=\delta_1-\overline \delta_1= C_1x$, with
 $
 C_1=[1,0,0,0].$ It is easy to verify that under this observation equation, the system is observable.
 On the other hand, the measurement of the real power $P_{12}=\beta \sin(\delta)=\beta \sin(\delta_1-\delta_2)$ can be represented as $y=C_2x$, with
 $
 C_2=[\beta\cos(\overline \delta),0,-\beta\cos(\overline \delta),0].
 $
 Different sensor systems affect observability, which characterizes if the measured values are sufficient to determine the interval (unmeasured) states.
 Under the system parameters in \emph{Example \ref{exm1}}, if we measure the power transfer $P_{12}$ with $y=C_2x$ the observability matrix is
\begin{align*}
W_2 = \begin{bmatrix}
     1 &         0 &  -1     &    0\\
         0  &  1 &         0  & -1 \\
 -329.5620 &  -0.2 & 329.5620 &   0.2067\\
   66.7912 & -329.5220 & -66.7912 & 329.5193
\end{bmatrix},
\end{align*}
which has rank $3$. As a result, this state space model is not observable in this case.

\end{rem}

\vspace{-9.5pt}
\subsection{Contingency Models}
\vspace{-2pt}

Power system contingencies are of diversified types. We list some of the common types.
\begin{enumerate}
\item \textbf{Transmission Line Grounding:}
A transmission line fault can change the impedance values $|Z|$ on the line.
For example, a single-phase grounding will reduce the impedance value. In contrast, high-impedance faults are very common in case of single-phase faults. Since line impedances are parameters in the matrix $A$, a line fault will cause a jump in the $A$ matrix value. \item \textbf{Transmission Line Breaking:} When a  transmission line breaks due to natural disasters or faulty components, the transmission line's impedance will experience a jump of $X_{ij}$ to a much bigger value;
\item \textbf{Generator Excitation System Fault:} Loss of excitation (LOE) is a common fault in generators. LOE causes a sudden decline of the terminal voltage $V$, with some other potential damages to the generator.
\item \textbf{Intentional Attack:} Cyber attackers may intentionally damage a sensor, a transmission line, a bus, creating a jump in system structure or parameters.
\end{enumerate}

Traditionally, contingencies are detected by special devices that monitor the targeted buses and lines. For example, impedance relays are very common devices for  protecting high-voltage transmission lines from faults. This paper introduces SHS models for contingency detection. Under this framework and our algorithms, a PMU on one bus can potentially detect impedance jumps of many lines without using special devices on these lines. This will be demonstrated in the case studies.

Power system contingencies can be generally modeled as jumps on system matrices. Mathematically, we list all scenarios of contingencies under study as a set $\clS=\{1,\ldots, m\}$ and use a jumping process $\alpha(t) \in \clS$ to represent the occurrence of the corresponding scenario. For example,
for the system in (\ref{Amatrix}), the above-listed faults are reflected on the coefficient $\beta= V_1 V_2/X_{12}$ as a switching of its value during contingency. If the excitation system for Bus $1$ experiences a loss-of-magnet fault on its excitor, then $V_1$ will drop. On the other hand, a partial transmission line fault, such as a one-phase line grounding,  changes the impedance value $X_{12}$, leading to a jump in $\beta$ value. Suppose that $\beta$ changes its value from $200$ to $100$. Then, the new $\overline\delta$ value is $\overline\delta_2=0.3047$ and the system's matrix becomes the following new one: 
 
  \begin{align*}
     A_2=\begin{bmatrix}
         0  &  1 &         0      &   0\\
 -190.7878  & -0.2 &  190.7878     &    0\\
         0    &     0  &        0  &   1 \\
  127.1919    &     0 & -127.1919  & -0.2067
\end{bmatrix}.
\end{align*}  
 
 
\subsection{Randomly Switched Linear Systems}
 
 The dependence of system matrices on contingencies can be represented by their values as functions of $\alpha$, expressed as $A(\alpha)$, $B_1(\alpha)$, $B_2(\alpha)$, $D_1(\alpha)$, $D_2(\alpha)$ and $C(\alpha)$.
Since contingencies occur randomly, $\alpha(t)$ is a stochastic process.
Including the jumping process into the system dynamics (\ref{total3}) introduces the following hybrid system:
    \beq{sys}
\!\!\! \!  \!\left\{ \! \!\!\begin{array}{rcl} \dot x \!\!\!\! \! \!&=& \!\!\!\!\! A(\alpha) x\!+\!B_1(\alpha) u\!+\!B_2(\alpha) u^n \! \!+\! D_1(\alpha) \zeta\!+\!D_2(\alpha) \zeta^n \!,\\
\!\!\!y \!\!\!\! \! \!& = & \! \!\!\!\!C(\alpha) x.
\end{array}\right.
\eeq
This system is an RSLS. The contingency detection problem aims to study joint discrete event detection and continuous state estimation of RSLSs.

The system matrices depend on the randomly switching process $\alpha(t)$ that takes $m$ possible values in a discrete state space $\clS=\{1,\ldots, m\}$. For each given value  $i\in \clS$, the corresponding linear time invariant (LTI) system in (\ref{sys}) with matrices $(C(i), A(i), B_1(i), B_2(i),D_1(i), D_2(i))$ is
called {\it the $i$th subsystem} of the RSLS. We introduce the following assumptions on the random switching process for the theoretical analysis.

\begin{asm}\label{asm1}
Given a sampling interval $\tau$, (i) the switching process $\alpha(t)$ can switch only at the  instants $k\tau$, $k=0,1,2,\ldots,$ that generates a stochastic
sequence $\{\alpha_k=\alpha(k\tau)\}$ ({\it the skeleton sequence}); (ii) The sequence $\{\alpha_k\}$ is independent and identically distributed  (i.i.d.) with probability ${\rm Pr}\{\alpha_k=i\}=p_i>0$, $i\in \clS$,
and $\sum_{i=1}^m p_i =1$; (iii) $\alpha_k$ is independent of $x(0)$ and the Brownian motion $w$.
\end{asm}

The main difference of \emph{Assumption \ref{asm1}} from \cite{cd1} is that $\alpha(t)$ cannot be directly measured and must be estimated here in this paper. Note also that power system management usually imposes certain intervals for data processing. For example,  PMU  data rate of the Power Xpert Meter is $1024$ samples per cycle. For contingency management, $160$ ms is the IEEE imposed limit for voltage sag/surge. For slower dynamics of power dispatch, a decision interval of $5$ minutes is commonly used in practice.
Mathematically, under this assumption, the random switching process can be treated as a discrete-time stochastic sequence, rather than a continuous-time process.

Under \emph{Assumption \ref{asm1}},
$A_k =A(\alpha_k)=\sum_{i=1}^m A(i) \one_{\{\alpha_k=i\}}$,
$B_k^1 =B_1(\alpha_k)=\sum_{i=1}^m B_1(i) \one_{\{\alpha_k=i\}}$, $B_k^2 =B_2(\alpha_k)=\sum_{i=1}^m B_2(i) \one_{\{\alpha_k=i\}}$,
 $C_k  =C(\alpha_k)=\sum_{i=1}^m C(i) \one_{\{\alpha_k=i\}}$,
where $\one_G$ is the indicator function of the event $G$: $\one_G =1$ if $G$ is true; and $\one_G =0$, otherwise. These are matrix-valued random variables. The sampled values of the signals are denoted by $x_k=x(k\tau)$, $y_k=y(k\tau)$.

The premise of this paper is to treat RSLSs whose initial states are unknown and whose switching sequence $\alpha_{k}$ cannot be directly measured. As a result, both the continuous state $x_k$ and discrete state $\alpha_k$ must be estimated from the known input $u(t)$ and the observed output $y(t)$. The available data set in a time interval  $[0,\tau)$ is given by
  the noise-free data set $\clD_{\tau}=\{y(t), t\in [0,\tau)\}$ for a given $\tau >0$.

\subsection{State Space Decomposition}

For the $i$th subsystem in $\clS$,  $A(i)$ and $C(i)$  are constant matrices, and its
 observability matrix
is 
  $W(i)=\left[\begin{array}{c} C(i)\\ C(i) A(i)\vspace{-3.2pt}\\ \vdots  \\ C(i) (A(i))^{n-1} \end{array} \right].$  
  Denote   $W_\clS= \left[\begin{array}{c} W(1)\vspace{-4pt}\\\vdots \\ W(m) \end{array} \right] $  
  as the combined observability matrix for $\clS$.
We note that both $W(i)$ and $W_\clS$ are deterministic matrices that contain only information on subsystems. They
do not involve actual switching sequences. Thus, they can be evaluated off-line.

\begin{asm}\label{asm2}
(i) Subsystems may be unobservable, namely, $\Rank (W(i))= n_i \leq n$, $i\in \clS$;
(ii) $W_\clS$ is full column rank.
\end{asm}

By \emph{Assumption \ref{asm2}}, since the $i$th subsystem
 may be unobservable, namely,  $\Rank(W(i))=n_i <n$, we construct
 $M_i=\Base(\ker(W(i))) \in \mathbb{R}^{n\times (n-n_i)}$ 
 and select any $N_i\in \mathbb{R}^{n\times n_i}$  such that
 $T_i=[M_i, N_i]$ 
 is invertible.
 The inverse of $T_i$ is decomposed into
 $T_i^{-1} =\begin{bmatrix}
     K_i\\ F_i
 \end{bmatrix} $,
 where \begin{small} $K_i\in \mathbb{R}^{(n-n_i)\times n}$\end{small} and   $F_i\in \mathbb{R}^{n_i\times n}$.

The state transformation $\wdt z^i =T_i^{-1} x$ can be decomposed into
$
 \wdt z^i =T_i^{-1} x
  =\begin{bmatrix} K_i x\\ F_i x \end{bmatrix} 
  = \begin{bmatrix} v^i \\  z^i \end{bmatrix} 
 $     
 where $ z^i \in \mathbb{R}^{n_i}$. Correspondingly, this coordinate transformation
 leads to the transformed matrices
 $  A^i = T_i^{-1} A(i) T_i$, $B^i=T_i^{-1} B(i)$, $ C^i = C(i) T_i$, with the structures
$A^i = \begin{bmatrix}
     A^i_{11} &  A^i_{12}\\
 0 & A^i_{22} 
 \end{bmatrix} $, $B^i= \begin{bmatrix} B^i_1\\ B^i_2 
 \end{bmatrix} $, $C^i= [0,  C^i_2]$     
 with $ A^i_{22} \in \mathbb{R}^{n_i \times n_i}$ and $ C^i_{2} \in \mathbb{R}^{1 \times n_i}$. As a result, if we focus only on the dynamics of the observable partial state $ z^i$, we have
  \beq{zi}
 \left\{ \begin{array}{rcl} \dot{z}^i &=&  A^i_{22}  z^i + B^i_2 u,\\
 y & =&   C^i_2  z^i, \end{array}\right.
 \eeq
 where $( C^i_2,  A^i_{22})$ is observable.

\section{Contingency Detection}\label{Design}

Under our stochastic hybrid system models, contingencies are represented by jumps in system structures and parameters. Mathematically, they are indexed by the stochastic process $\alpha_k$. Consequently, contingency detection in power systems becomes a problem of estimating $\alpha_k$ correctly when it jumps, on the basis of output observations. At $t=k\tau$, the internal continuous state $x_k=x(k\tau)$ is also unknown and must be estimated from the same output $y$. As a result, we must develop reliable joint estimation algorithms for estimating both $\alpha_k$ and $x_k$ simultaneously.

This joint estimation problem in power systems encounters many challenging issues, including detectability, joint estimation algorithms, convergence, and reliability. For instance, power systems are complicated network systems. In the $N-1$ reliability standard of power systems, one considers a fault on one transmission line, with other transmission lines under normal operating conditions. It will be shown in our case studies that the resulting system matrices $A(i)$ will typically share common eigenvalues since only a small part of the grid has changed its parameter values. In our recent theoretical work \cite{WY2}, it has been shown that without using input assistance, the stochastic hybrid system is not detectable, namely, some contingencies cannot be detected by the existing sensor systems.

Due to this complication, to ensure the ability to detect contingencies, it is necessary to add a small probing input $u$.
When the input is applied to an unknown subsystem with unknown initial state, the output contains both the input response and initial-state response. Input design and contingency detection algorithms are critical in this complicated situation.

\subsection{Input Design Principles}

First, we demonstrate by a simple example that the input must be suitably designed. Otherwise, even with input assistance, contingency detection may not be achievable.
 \begin{exm}\label{exm3} {\rm
 Consider an RSLS with two subsystems:   $A(1) =\begin{bmatrix}-4 \!\!&\!\! 0\\
0 \!\!&\!\!  -5\end{bmatrix}$, $B(1)=\begin{bmatrix}1\\
1 \end{bmatrix}$, $C(1)=
 [1 \ 1]$; $A(2) =\begin{bmatrix}-4 \!\!&\!\!  0\\
0 \!\!&\!\!  -10 \end{bmatrix}$, $B(2)=\begin{bmatrix} 1\\
1 \end{bmatrix}$, $C(2)=[1 \ 2]$.
We need to detect $\alpha_k \in \{1,2\}$ by using the output measurement data.

We first note that these two subsystems have the same eigenvalue $-4$, so they cannot be distinguished without assistance from a probing input. Suppose that we use the unit step $U(s)=1/s$ as the input signal. Then the two subsystems have respective transfer functions
$ G_1(s)= {1\over s+4} + {1\over s+5}={2s+9\over (s+4)(s+5)}\in \clR$;
$G_2(s)= {1\over s+4} + {2\over s+10}={3s+18\over (s+4)(s+10)}\in \clR.
$
\vspace{1pt} Their  total respective
responses to the input and (unknown) initial state are
\bea
\!\!y_1(t)\ad = a_1 e^{-4t} + a_2 e^{-5t} + 9/20-(1/4)e^{-4t}-(1/5)e^{-5t}, \\
\!\!y_2(t)\ad = b_1 e^{-4t} + b_2 e^{-10t} + 9/20-(1/4)e^{-4t}-(1/5)e^{-10t},
\eea
for $t\in [0,\tau)$,
where $a_1, a_2, b_1,b_2$ are determined by the initial states.
Denote their difference as
\bea
\delta(t)\ad = y_1(t)-y_2(t)\\
\ad = (a_1-b_1) e^{-4t} + (a_2-1/5)e^{-5t}+(b_2-1/5)e^{-10t}.
\eea
Then, the difference becomes $\delta(t)\equiv 0, t\in [0,\tau)$ if $a_1=b_1$, $a_2=1/5$, $b_2=1/5$.
In other words, we cannot uniquely determine if the subsystem is $\alpha_k=1$ or $\alpha_k=2$ in this case.}
\end{exm}

The theoretical foundation of this method was introduced in \cite{WY2} and is summarized below. As a first-time introduction of this method into power system contingency detection, some algorithm details are added and important related features of power system dynamic models and their impact on contingency detection are highlighted in the following part.

\emph{Example \ref{exm3}} indicates that the input signal must be suitably designed to enhance detectability on contingencies, and some design principles must be followed in selecting suitable inputs for contingency detection.
Consider the set $G=\{G_i, i=1,\ldots,m\}$ of $m$ distinct subsystems. The set of poles of $G_i$ (or equivalently the eigenvalues of $A(i)$) is $\Lambda_i$ and $\Lambda= \cup_{i=1}^m \Lambda_i$.

\begin{asm}\label{asm8}
Let $\clU\subset \clR_0$ be the set of non-vanishing inputs $u$ whose Laplace transforms $U(s)$ satisfy the following conditions:
(i) $U(s)={b(s)\over a(s)}$ is coprime, namely, no common pole-zero pairs (i.e., no pole-zero cancellation);
(ii) $U(s)$ contains at least one pole $\lambda$ of any multiplicity $q\geq 1$ such that  $\lambda\not\in \Lambda$ and $G_i(\lambda)$, $i=1,\ldots,m$, are distinct.
\end{asm}

The following result from \cite{WY2} forms the foundation for the input design.

\begin{thm}\label{thm3}\cite{WY2}
For the set of distinct subsystems $G=\{G_i, i=1,\ldots,m\}$, if the input $u\in \clU$, where $\clU$ is given in  \emph{Assumption \ref{asm8}}, then for any $\tau>0$, the true subsystem can be uniquely determined from the data set $\clD_\tau=\{y(t)\not\equiv 0, t\in [0,\tau)\}$, regardless of the actual initial state $x(0)$.
\end{thm}

\vspace{-12pt}
\subsection{Two-Time-Scale Framework and Joint Estimation Algorithms}

The contingency detection and continuous-state observers will be implemented in a two-time-scale
framework. Each time segment $[k\tau, (k + 1)\tau)$ is divided into
two intervals. The first smaller interval $[k\tau, k\tau + \tau_{0}]$
is designated for estimating $\alpha_k$ (that is, identifying the active
subsystem). During this time interval, the probing input $u$ that satisfies the conditions of \emph{Theorem \ref{thm3}} is applied to assist
in determination of $\alpha_k$. Once $\alpha_k = i$ is correctly estimated, in the second interval
$[k\tau +\tau_0, (k+1)\tau)$, a feedback-based observer is implemented
for the $i$th subsystem to estimate its observable sub-state
$z^{i}$.

\subsubsection{Detection of  $\alpha_k$ using Data in $[k\tau,k\tau+\tau_0]$}

$\alpha_k$ is detected by the following algorithm. 
\begin{algorithm}
	\caption{Detection of $\alpha_k$ under Unknown Initial State}
	
	\begin{algorithmic}[1]

 \State Calculate  the eigenvalues of all $A(i)$ and denote as $\Lambda$. Take  $u$ satisfying  Assumption \ref{asm8} as the designed input.

        \State Under the designed input $u$, collect and sample the output data on $y(t)$ in $[k\tau,k\tau+\tau_0]$.
        Define a small sampling interval $t_s$: Let $N_0=\tau_0/t_s$ be an integer.
        Obtain sampled values $y(k\tau+\ell t_s)$, $\ell = 0, \ldots, N_0$.

		\State Compute the input responses of the subsystems (assuming zero initial condition):
		$
		y^{input}_i(\ell)=(G_i u)(k\tau+\ell t_s)$, $\ell = 0, \ldots, N_0$. 
        Since the input $u$ and the system transfer functions $G_i$ are known in advance, these responses can be computed offline and stored.
		
        \State Derive the net initial state responses of the subsystems:
        $
		y^{net}_i(\ell)=y(k\tau+\ell t_s)-y^{input}_i(\ell)$, $\ell = 0, \ldots, N_0$. 

        \State Estimating the initial observable sub-states of the subsystems:
        Derive the numerical Gramians of the subsystems
        $\Gamma_i= \sum_{\ell=0}^{N_0} t_s e^{A^{\top}(i)\ell t_s}C^{\top}(i)C(i)e^{A(i)\ell t_s}$ and
        $Y_i=\sum_{\ell=0}^{N_0} t_s e^{A^{\top}(i)\ell t_s}C^{\top}(i)y^{net}_i(\ell)$. 
        Then,
        $ \wdh x^i_k= \Gamma_i^{-1} Y_i$. 

		\State Calculate the total estimated outputs of the subsystems:
        $\wdh y_i(\ell)= C(i)e^{A(i)\ell t_s}\wdh x^i(k\tau)+ y^{input}_i(\ell).$ 

        \State Calculate the output prediction errors of the subsystems
		$
		\varepsilon_i = {1\over N_0+1}\sum_{\ell=0}^{N_0} |\wdh y_i(\ell)-y(k\tau+\ell t_s)|$. 
        This error measure may be replaced by the common Euclidean norm or the $\max$ norm.
		
		\State Determine $\alpha_k$:
		$
		\wdh \alpha_k = \arg\min_{i=1,\ldots, m} \varepsilon_i.
		$
	\end{algorithmic}
\end{algorithm}

\subsubsection{Observer Design for $x$ in $[k\tau+\tau_0, (k+1)\tau)$}

After determining $\alpha_k=i$ correctly, an observer can be designed to estimate $z^i(k\tau+\tau_0)$. The errors in estimating $z^i$ and $z$ are  denoted by $e_i= z^i-\wdh z^i$ and $e= z-\wdh z$, respectively. Denote
\begin{small}
   $
\mu^i(t) =\|e_i(t)\|, \mu^i_k =\|e_i(k\tau)\|, \mu(t)= \|e(t)\|, \mu_k =\|e(k\tau)\|.
$ 
\end{small}

\begin{asm}\label{asm6}
		We assume that (i) The RSLS  has  independent subspace error dynamics, namely
	$\dot z^i$ depends on $z^i$ only, independent of $\alpha_k$, under zero input.
	For such systems, the {\it subsystem  state equation} will be
	$ \dot e_i =  A^i_{22} e_i$
	in open loop without input; (ii) $B_k$ is known.
\end{asm}

Under \emph{Assumption \ref{asm6}}, we consider the following three cases in the error analysis:

\textbf{Case 1: $t\in [k\tau,k\tau+\tau_0]$}

In this time interval, all subsystem observers are running open-loop. Since a probing input is applied, under \emph{Assumption \ref{asm6}}, the dynamics of $z^i$ are
$
\dot z^i = F_i \dot x
= F_i A_k x + F_i B_k u
=  A^i_{22} z^i + F_i B_k u.
$
The observer is
$
\dot{\wdh{z}^i}=  A^i_{22} \wdh z^i + F_i B_k u.
$
It follows that the error dynamics are
$ \dot e_i =  A^i_{22} e_i$, and
$
\|e_i(k\tau+\tau_0)\| \leq \gamma_0^i,
$
for some $\gamma_0^i > 0$. Let $\gamma_0 = \max_{i=1,\ldots, m} \gamma_0^i$.

\textbf{Case 2: $t\in [k\tau+\tau_0,(k+1)\tau)$ and $\alpha_k\neq i$}

In the interval $t\in [k\tau+\tau_0,(k+1)\tau)$, the input $u\equiv 0$.
When the $i$th subsystem is running open loop, we have the error bound
$
\mu^i_{k+1} \leq \gamma^i_1 \|e_i(k\tau+\tau_0)\| \leq  \gamma^i_1\gamma^i_0 \mu^i_k, \alpha_k\neq i,
$
for some constant $\gamma^i_1$.
Let
$\gamma_1=\max_{i=1,\ldots, m} \gamma^i_1.$

\textbf{Case 3: $t\in [k\tau+\tau_0,(k+1)\tau)$ and $\alpha_k= i$}

Observe that
if $\alpha_k=i$, the observer error dynamics for the $i$th subsystem are
$
\dot e_i = ( A^i_{22}-  L_i   C^i_2) e_i = A^i_c e_i.
$
By designing the observer gain $L_i$ properly,
$A^i_c= A^i_{22}-  L_i   C^i_2$
can have $n_i$ eigenvalues with real part less than $-a_i$ with $a_i>0$. Under the given $\tau$,
for some $c>0$,
$ \| e^{A^i_c \tau}\| \leq c e^{-a_i \tau}$
which can be made arbitrarily small by choosing sufficiently large $a_i$.
Consequently,
$
\mu^i_{k+1} \leq \gamma^i_c  \|e_i(k\tau+\tau_0)\|
\leq \gamma^i_c \gamma_0^i \mu^i_k,
$
where $\gamma^i_c$  can be made arbitrarily small.
Denote
$\gamma_c=\max_{i=1,\dots,m}\gamma^i_c.$
The actual value $\gamma_c$ will be selected later to ensure convergence of the organized observer for the entire system.

In summary, combining the three cases,
we have \vspace{-2pt} 
\begin{small}
\begin{align*} 
	\dot e_i=\left\{\!\begin{array}{ll} A^i_{22} e_i, & t\in [k\tau, k\tau+\tau_0], \\
	I_{\{\alpha_k=i\}} A^i_c e_i+ I_{\{\alpha_k\neq i\}} A^i_{22} e_i, & t\in [k\tau+\tau_0, (k+1)\tau).
	\end{array}\right.
\end{align*}\vspace{-2pt} 
\end{small}
It follows that the errors are bounded by
$
\mu^i_{k+1}  \leq \gamma^i_k \mu^i_k,
$
with
$
\gamma^i_k=I_{\{\alpha_k=i\}}\gamma^i_c\gamma_0^i  +I_{\{\alpha_k\neq i\}} \gamma_1^i\gamma^i_0.
$
Consequently,
$\mu^i_k\leq \left(\Pi_{j=1}^k \gamma^i_j\right) \mu^i_0.$ Under \emph{Assumption \ref{asm1}}, the process $\{\gamma^i_k\}$ is i.i.d. with
$
P(\gamma^i_k=\gamma^i_c\gamma_0^i)=p_i, P(\gamma^i_k=\gamma_1^i\gamma^i_0)= 1-p_i.
$
The following result can be easily obtained and we omit the proof here. Subsystem observers are designed to satisfy (\ref{maxg}).

\begin{lem}\label{lem5}
	Under \emph{Assumptions \ref{asm1}} and \emph{\ref{asm6}}, for any $\gamma_* < 1$, the pole positions in the observer design can be selected
	such that\vspace{-4pt}
	\beq{maxg} \gamma^i=(\gamma^i_c\gamma_o^i)^{p_i} (\gamma^i_1\gamma_o^i)^{(1-p_i)}\leq \gamma_* <1.\eeq
\end{lem}\vspace{-5pt}

\vspace{-13pt}
\subsection{Convergence Analysis}

\begin{asm}\label{asm7}
	$\alpha(t)$  is independent of $e_i(0)$.
\end{asm}

Define the  continuous-time error $\mu^i(t)=\|e_i(t)\|$, which is a scalar stochastic process. Also, define $e(t)=[ e_1(t), \dots, e_m(t)]^{\top}$. The estimation error on $x$ is $\e(t)=x(t)-\wdh x(t)$. Therefore, we can obtain the convergence results.

\begin{thm}\cite{WY2}
	Under \emph{Assumption \ref{asm7}} and the observer design in \emph{Lemma \ref{lem5}}, we have (i) $\mu^i_k$ converges strongly and exponentially to $0$ as $k\to \infty$; (ii) $\mu^i(t)$ converges strongly and exponentially to $0$ as $t\to \infty$; (iii) $\|\e(t)\|$ converges strongly and exponentially to $0$ as $t\to \infty$.
\end{thm}


\section{Case Studies}\label{case}

\subsection{IEEE 5-Bus system}
\begin{figure}[htbp] \vspace{-10pt}
	\setlength{\unitlength}{0.1 in}
	\begin{center}
		\includegraphics[height=2.7cm]{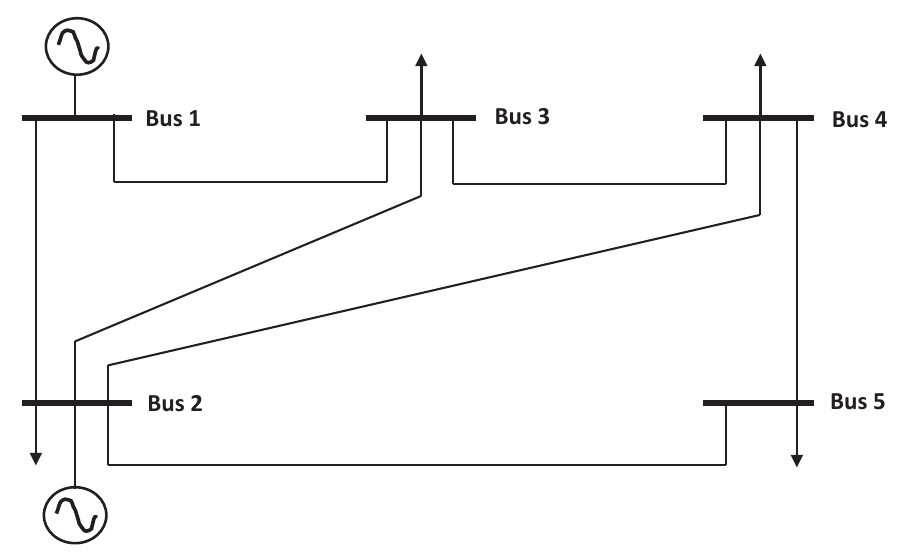} \vspace{-8pt}
	    \caption{IEEE 5-Bus System}
 \label{5bus} 
	\end{center}
\end{figure}

\vspace{-3pt}
In this subsection, we use the IEEE 5-Bus system, shown in Fig. \ref{5bus}, to illustrate model derivations, the design process, performance evaluation, and related issues. The power system structure and data are from the open-source information in \cite{Tan}. Bus 1 and Bus 2 are dynamic dispatchable  buses and Buses 3-5 are non-dynamic non-dispatchable  buses.   

In the original system, Bus 1 is a slack bus with unlimited and instantaneous power and its voltage (both magnitude and angle) is a reference point. This will significantly simplify the SHS model. In consideration of renewal generation situations under potential islanding operations (all bus angles can change dynamically during operation and no bus has unlimited power), we consider the more general scenarios and  designate both Buses 1 and 2 as PV buses with bus voltage magnitudes controlled to their rated values.
 In light of rapid advancement in Var compensation technology such as flexible AC transmission systems (FACTS), we assume that all buses have their voltage magnitudes maintained near the rated values during normal operations, but their values may jump during contingency.

\subsubsection{Stochastic Hybrid System Models}
$$ $$
\vspace{-30pt}

\textbf{(i) Dynamic Systems}

The most common dynamic generator types are synchronous generators \cite{K}. Denote
$
\omega_1=\dot \delta_1, z_1^d=[\delta_1; \omega_1], \omega_2=\dot \delta_2, z_2^d=[\delta_2; \omega_2].
$
The dynamic systems are
$ M_1 \dot \omega_1 + g_1(\omega_1) = P_1^{in} - P_1^L+ P_1^{21}+P_1^{31},
M_2 \dot \omega_2 + g_2(\omega_2) = P_2^{in} - P_2^L+ P_2^{12}+P_2^{32}+ P_2^{42}+P_2^{52},
$
where  the real power flow from Bus $i$ to Bus $j$ is
$P_j^{ij}= {V_j^2\over X_{ij}} \cos( \theta_{ij}) - {V_iV_j\over X_{ij}}\cos (\theta_{ij}+\delta_{ij}),$
and $\delta_{ij} =\delta_i-\delta_j$. The damping term $g_i(w_i)$ has the linear part $b_i \omega_1$ with $b_i>0$, $i=1,2$. The three non-dynamic non-dispatchable  buses have real-power equations
$P_3^L  = P_3^{13}+P_3^{23}+ P_3^{43}, P_4^L  = P_4^{24}+P_4^{34}+ P_4^{54}, P_5^L  = P_5^{25}+P_5^{45}$.
Denote
$
z^{d}=[z_1^d; z_2^d], z^{nd}=[\delta_3; \delta_4; \delta_5]$. The dynamic systems  can be expressed as a nonlinear state equation
$
\dot z^d =  F^{0}(z^d, z^{nd}) + B_1 v+D_1 \ell^{d}
=  F(z^d, \ell^{nd}) + B_1 v+D_1 \ell^{d},
$
where
$
v=\left[P_1^{in};
P_2^{in} \right], \ell^{d}=\left[P_1^L;P_2^L\right],\ell^{nd}=\left[P_3^L;
P_4^L;P_5^L \right],$ and
\begin{small}\begin{align*}
B_1=\begin{bmatrix}
0&0  \\
1/M_1 &0 \\
0&0  \\
0&1/M_2 
\end{bmatrix}, ~~D_1=\begin{bmatrix} 0&0  \\
-1/M_1 &0 \\
0&0  \\
0&-1/M_2
\end{bmatrix}.
\end{align*}\end{small}\vspace{-2pt} 
Denote the perturbations from the nominal values as
$x\!=\!z^{d}\!-\!\overline z^{d}, u\!=\!v\!-\!\overline v, \zeta\!=\!\ell^{d}\!-\!\overline \ell^{d}, \zeta^n\!=\!\ell^{nd}\!-\!\overline \ell^{nd}$. 
By (\ref{total2}), the dynamic systems can be linearized near the nominal operating points as
$
\dot x \!= \!A x\!+\!B_1 u\!+\! D_1 \zeta \!+\! D_2 \zeta^{n},
$
where the matrices are  the related Jacobian matrices
$
A = \left.{\partial F(z^d, \ell^{nd})\over \partial z^d}\right|_{\tiny{\begin{array}{c} z^d=\ol z^d, \ell^{nd}=\ol \ell^{nd}\end{array}}},
D_2 = \left.{\partial F(z^d, \ell^{nd})\over \partial \ell^{nd}}\right|_{\tiny{\begin{array}{c} z^d=\ol z^d, \ell^{nd}=\ol \ell^{nd}\end{array}}}.
$

The nominal operating condition defined in \cite{Tan,Simudata}  is used here with the nominal bus voltages, generation powers and load powers listed in Table \ref{buses} with real power  $P$ (MW) and reactive power $Q$ (MVar). The base MVA is $S_B= 100$ MVA  and the base voltage is $V_B= 230$ kV.
The bus line parameters, shown in Table \ref{lines}, are extracted from \cite{Tan}.
\begin{table}[h!] 
\centering
\caption{IEEE 5-Bus System Bus Data}   \vspace{-5pt}
\label{buses}
\begin{tabular}{|c|c|c|c|c|c|} \hline
Bus  & $V$ (p.u. $\angle$ rad)  &  $P$  & $Q$  &  $P_L$  & $Q_L$   \\ \hline
1 & $1.06 \angle 0$ & $129$ & $-7.42$ &  $0$ & $0$ \\ \hline
2 & $1.0474 \angle -2.8063$  & $40$ & $30$ & $20$ & $10$ \\ \hline
3  & $1.0242 \angle -4.997$   & $0$ & $0$ & $45$ & $15$  \\ \hline
4  & $1.0236 \angle -5.3291$   & $0$ & $0$ & $40$ & $5$ \\ \hline
5 & $1.0179\angle -6.1503$   & $0$ & $0$ & $60$ & $10$  \\ \hline
\end{tabular}
\end{table}  
\begin{table}[h!]  
\centering
\caption{IEEE 5-Bus System Line Parameters}  \vspace{-5pt}
\label{lines}
\begin{tabular}{|c|c|c|c|} \hline
 Line &  Resistance (p.u.)   &  Reactance (p.u.) & Z (p.u $X\angle \theta $ rad)   \\ \hline
1-2 & 0.02  & 0.06 & $0.06   \angle 1.25$ \\ \hline
1-3 & 0.08  & 0.24 & $0.25  \angle 1.25$ \\ \hline
2-3  & 0.06  &  0.25 & $0.26 \angle  1.33$ \\ \hline
2-4  & 0.06  & 0.18 & $0.19  \angle 1.25$ \\ \hline
2-5 & 0.04  & 0.12 & $0.13 \angle  1.25$ \\ \hline
3-4 & 0.01  & 0.03 & $0.03 \angle  1.25$ \\ \hline
4-5 &  0.08  &  0.24 & $0.25\angle  1.25$ \\ \hline
\end{tabular}  
\end{table}

Under the per unit system, the normalized generator parameters are $M_1=1.9$ and  $b_1=0.2$ with equivalent time constant $T_1=M_1/b_1=9.5$ second for Generator 1, and  $M_2=0.9$, $b_1=0.16$ with equivalent time constant $T_2=M_2/b_2=5.625$ second for Generator 2. Under the aforementioned operating conditions, we obtain  \vspace{-3pt}
\begin{align*}\begin{small}
A=\begin{bmatrix}
        0  &   1 &         0   &      0\\
    7.7926 &  -0.1053 &  -7.7926   &      0\\
         0     &    0     &    0   & 1\\
  -20.3866     &    0  & 20.3866  & -0.1778
\end{bmatrix} .
\end{small}
\end{align*}

\textbf{(ii) Sensor Systems}

This paper aims to present a framework in which a small number of sensors can potentially detect a large set of contingencies. For example, if a power system has $50$ buses of which $10$ buses are dynamic and others are non-dynamic. If each dynamic bus has a second-order state space model, then the virtual power grid model will be of order $20$. It is noted that for cost reduction and maintenance simplification, it is highly desirable to reduce sensor complexity. Then a related question is: Will it be possible to use only one PMU to achieve one-line fault detection($N-1$ scenario) on all lines? Our algorithms indicate that this is possible, as long as the transfer functions from the control inputs to the sensor are distinct and the input is properly designed.
  In this simulation study,  we will use the sensor that measures $\delta_1$ (a PMU), i.e., $C=[1,0,0,0].$ We will demonstrate that although this is a voltage phasor sensor, it is sufficient for detecting a line fault.

\subsubsection{Evaluation Scenarios and Input Design}

In consideration of the N-1 reliability requirements in power systems, we focus on a fault on one transmission line with different scenarios for evaluation. Line $(2,3)$  which is the longest transmission line in the system,  is selected. The line faults are characterized by jumps in the impedance values $X_{23}$.
Four cases are considered: (1) Normal Operation: $\alpha=1: X_{23}=0.26$, with probability $p_1=0.9$;
(2) Line Fault 1 (Reduced Impedance): $\alpha=2: X_{23}=0.1$, with probability $p_2=0.06$;
(3) Line Fault 2 (single line to ground fault): $\alpha=3: X_{23}=0.06$, with probability $p_3=0.03$;
(4) Line Fault 3 (Disconnection): $\alpha=4: X_{23}=10000$, with probability $p_4=0.01$.

The corresponding $A$ matrices are :
\begin{small}
 \begin{align}
&A ( 1 )  =   \begin{bmatrix} 
      0      &       1      &              0        &     0\\
    7.7926  &  -0.1053  &   -7.7926    &    0\\
         0     &    0     &    0    &  1\\
  -20.3866      &     0   &  20.3866   & -0.1778
\end{bmatrix}   ,\nonumber\\
& A(2)=\begin{bmatrix}
0  &    1 &          0   &       0\\
7.7967  &  -0.1053 &   -7.7967      &    0\\
 0      &    0     &     0  &   1\\
 -20.5843  &   0 &  20.5843 &  -0.1778
\end{bmatrix} ,\nonumber\\
&A(3)=\begin{bmatrix}
0  &    1 &          0   &       0\\
7.7978 &   -0.1053&    -7.7978   &       0\\
         0  &        0     &     0    & 1\\
  -20.6409   &       0  &  20.6409  &  -0.1778
\end{bmatrix} ,\nonumber\\
&A(4)=\begin{bmatrix}
0  &    1 &          0   &       0\\
    7.7571 &   -0.1053 &   -7.7571  &        0\\
         0   &       0     &     0   &  1\\
  -18.6540    &      0 &   18.6540 &   -0.1778
\end{bmatrix}  ,\nonumber
\end{align}
\end{small}
respectively. It is easy to verify that under $C=[1,0,0,0]$, the corresponding observability matrices $W(1)$, $W(2)$, $W(3)$, $W(4)$ are full rank. As a result, the stochastic hybrid system has observable subsystems. For this reason, state decomposition is not needed, and we will directly estimate $x$ under each detected subsystem.

We first calculate the eigenvalues of $A(i)$ and obtain their eigenvalues,
which are \begin{small}
    $\{ -5.388, 5.2302,0,   -0.1253\},$ $ \{  -5.407,   5.2491,0,   -0.1252\}$, $\{  -5.412,    5.2545,0,  -0.1251\}$, $\{   -5.2181,   5.0616,0,-0.1266\}$
\end{small}. Since they share the common eigenvalue $0,$ an input is needed to detect different systems.
Select  $u=a\sin t$ with $a>0$,  which satisfies the conditions of Theorem \ref{thm3} (namely, $U(s)=\frac{a}{s^2+1}$, Assumption \ref{asm8} is satisfied).
By using a small $a$, this probing signal will have a negligible impact on the system's normal operation. For this case study, $a=0.1$ is used, although smaller values of $a$ can still work.

\subsubsection{Results and Discussions}

$$ $$
\vspace{-30pt}

\textbf{(i) Detection of Subsystems}

\begin{figure}[htbp] \vspace{-10pt}
	\setlength{\unitlength}{0.1 in}
	\begin{center}
		\includegraphics[height=4.6cm]{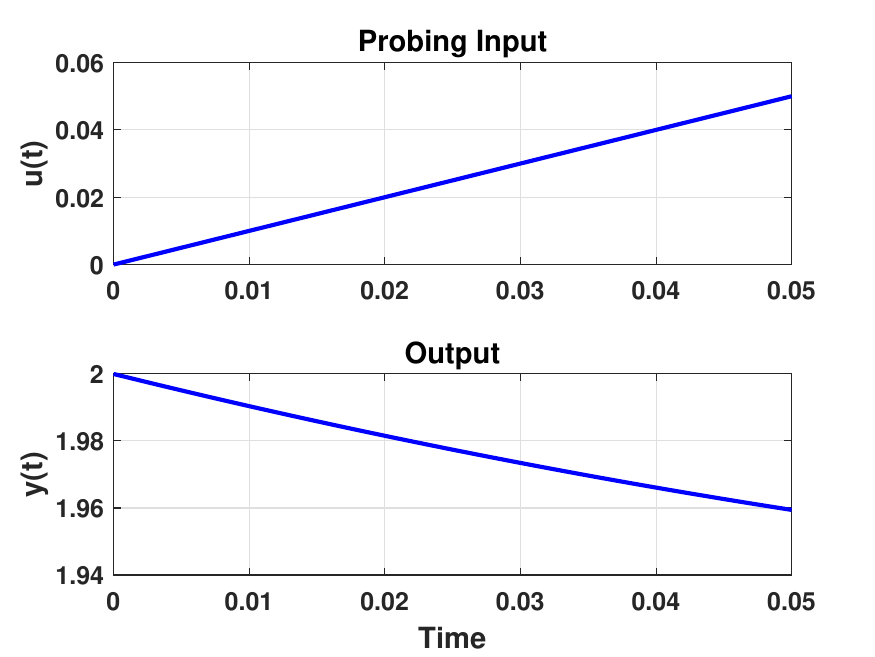} \vspace{-8pt}
	    \caption{The probing input $u(t)$ and the output $y(t)$ in $[0,\tau_0)$.}
    \label{fig0}
	\end{center}
\end{figure}

As an example, taking $\tau = 2.5$ and $\tau_0=0.05$, we show the detection of $\alpha_0$ for $t\in [0, \tau_0)$. Suppose the initial state is $[2,-1,1,2]$, and the true $\alpha_0=1.$ Now we estimate $\alpha_0$ by Algorithm 1. Under the probing input $u(t)=0.1\sin(t)$, Fig. \ref{fig0} shows the  curve of $u(t)$ and $y(t)$. According to Algorithm 1, we calculate the output prediction errors of the four subsystems, then we obtain $\varepsilon_1 = 2.7464\times 10^{-14},\ \varepsilon_2 = 1.6518\times 10^{-9},\ \varepsilon_3 =2.1253\times 10^{-9} ,\ \varepsilon_4 =1.4480\times 10^{-8}.$ Therefore, $\wdh{\alpha}_0=1$, which detects the subsystem  accurately.

\textbf{(ii) Observer Design and Convergence}

The pole placement design is used for designing observer feedback gains.
For example, if we choose the desired closed-loop poles as $\lambda=[-4.8,-3.6,-4,-4.4]$, then the Matlab function $L_i= place(A^{\top}(i),C^{\top}, \lambda)$, $i=1,2,3,4$, yields the suitable feedback gains and the closed-loop error dynamics with
$A_c^i=A(i)-L_i C$, $i=1,2,3,4$.

\begin{figure}[htbp] \vspace{-15pt}
	\setlength{\unitlength}{0.1 in}
	\begin{center}
		\includegraphics[height=5.5cm]{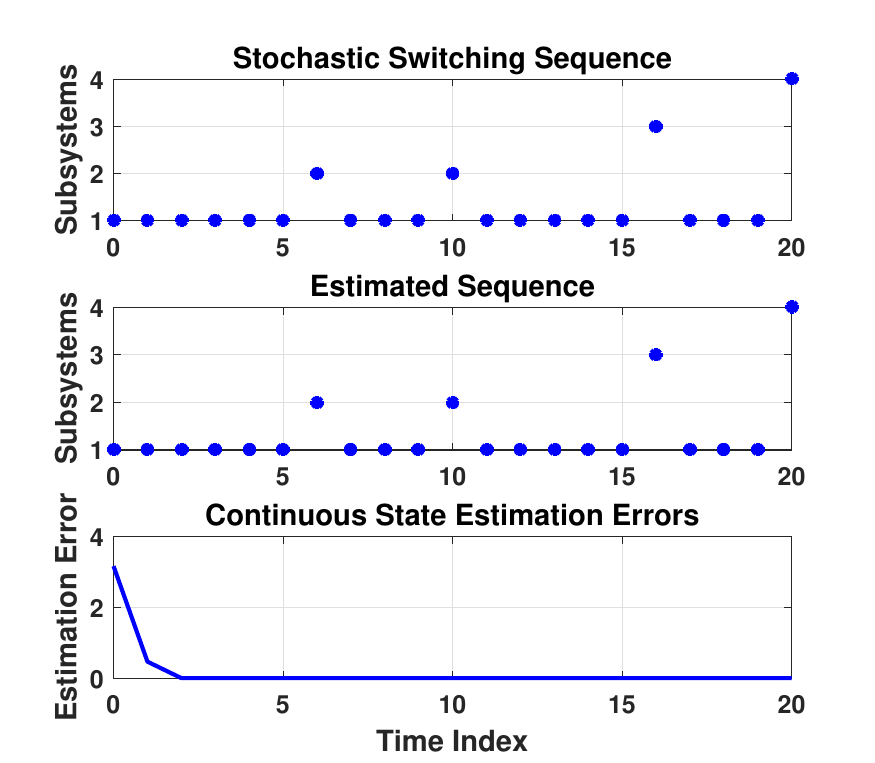} \vspace{-8pt}
	    \caption{The detection of $\alpha_k$ and the estimation error trajectory.}
    \label{fig1}
	\end{center}
\end{figure}

Take $\tau = 2.5$ and $\tau_0=0.05$.
The initial estimation error is selected to be $e(0)=[2,-1,1,2]^{\top}$ with the error norm $\sqrt{10}$.
Fig. \ref{fig1} shows that $\alpha_k$ can be accurately detected and the estimation error is convergent.

\textbf{(iii) Robustness against Measurement Errors}

\begin{figure}[htb] \vspace{-15pt}
	\setlength{\unitlength}{0.1 in}
	\begin{center}
		\includegraphics[height=5.5cm]{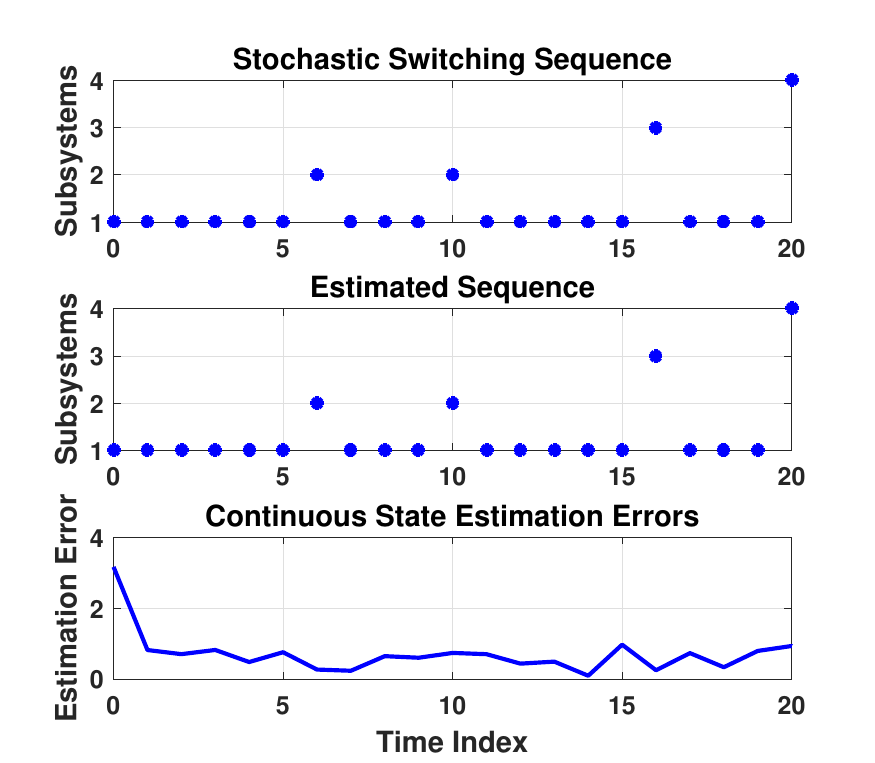} \vspace{-8pt}
	    \caption{The detection of $\alpha_k$ and the estimation error trajectory.}
		\label{fig6}
	\end{center}
\end{figure}

We now consider measurement noise and show the impact
of output measurement errors on contingency detection accuracy.
For the one-sensor case (i.e., $C=[1,0,0,0]$), suppose that the standard deviation is
$\sigma=0.005$, and the measured output value is $\hat{y}=y+\sigma d$, where $d$ is the noise  with uniform distribution in $[-0.5,0.5]$.
Take $\tau=2.5,\tau_0=0.05,\lambda=[-4.8,-3.6,-4,-4.4]$.
Fig. \ref{fig6} shows that $\alpha_k$ can still be accurately detected. However,  the steady-state error is big.

\begin{figure}[htb] \vspace{-15pt}
	\setlength{\unitlength}{0.1 in}
	\begin{center}
		\includegraphics[height=5.5cm]{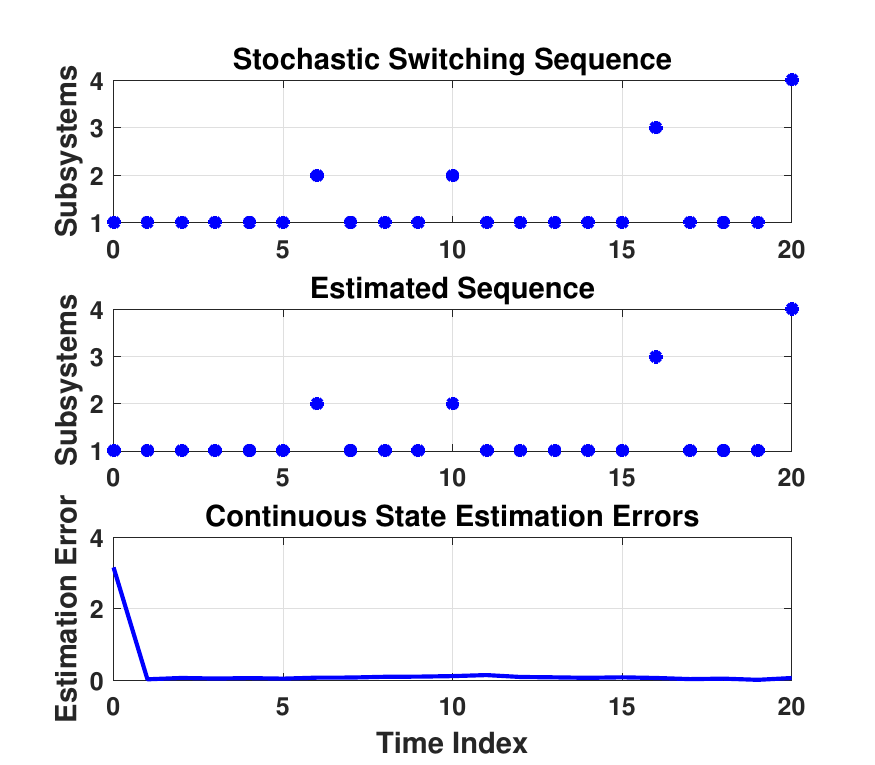} \vspace{-8pt}
	    \caption{The detection of $\alpha_k$ and the estimation error trajectory.}
		\label{fig4}
	\end{center}
\end{figure}

If we use two sensors $\delta_1$ and $\delta_2$, i.e., $C=\begin{bmatrix}
    1 \!&\! 0 \!&\! 0\!&\!  0\\
    0\!&\!  1 \!&\! 0\!&\!  0
\end{bmatrix}$. Suppose $\sigma=0.005 I_2$, then under the same $\tau$ and $\tau_0$, and the same pole positions, the  steady-state error becomes smaller, see Fig. \ref{fig4}.

\subsection{IEEE 33-Bus system }\label{case2}

To elucidate our algorithms in a more comprehensive context, we utilize the IEEE 33-bus system \cite{Wu, Wong} as an illustrative example in our research.

\subsubsection{ Modeling and Linearization of the IEEE 33-Bus System}

\begin{figure}[htbp] \vspace{-12pt}
	\setlength{\unitlength}{0.1 in}
	\begin{center}
		\includegraphics[height=4cm]{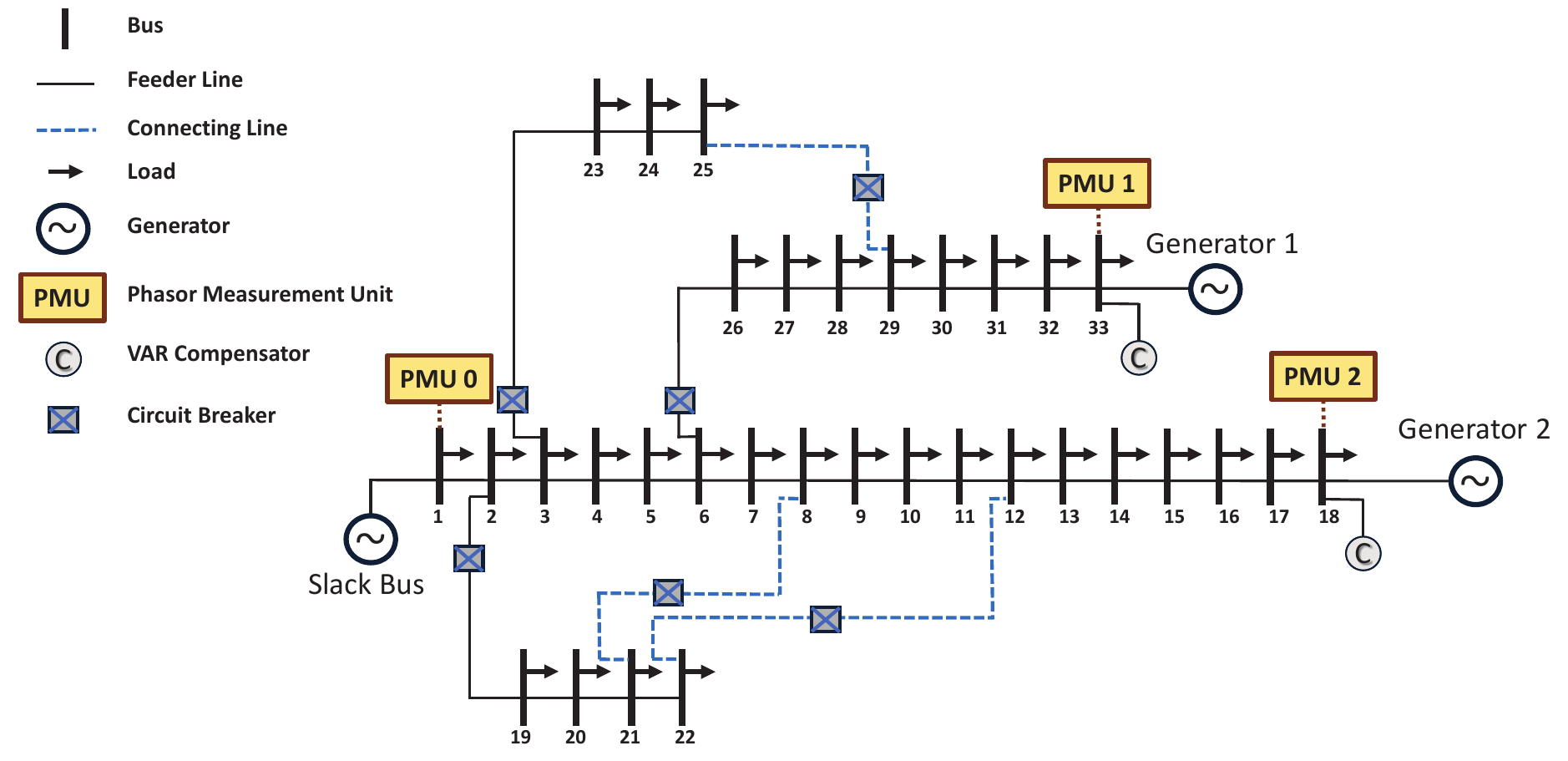}\vspace{-8pt}
	 \caption{The enhanced IEEE 33-bus distribution test system.}
\label{33bussystem}
	\end{center}
\end{figure}

The original 33-bus system \cite{Wu} contains one slack bus tied to the large grid and the rest buses are PQ-type load buses. For evaluation of renewable systems, more local generators have been added. Following the enhanced 33-bus evaluation system proposed in \cite{enhanced33bus}, in this simulation study, Bus 1 remains as a slack bus and two generators are added, at  Bus 18 and Bus 33,  shown in Fig. \ref{33bussystem}. The generator buses are dynamic buses whose local state space models for real power management are represented by their swing equations. All other buses remain as PQ-type load buses as in the original configuration and non-dynamic. The slack bus voltage is set as the reference bus with constant voltage  $1\angle 0$ (pu), whose $P$ and $Q$ injections are unlimited and instantaneous in balancing powers in each step. Consequently, the slack bus is non-dynamic. All bus and load parameters are from the power flow data in \cite{Wu} and obtained from the 33-bus case file  in MATPOWER \cite{Wong, MATPOWER1, MATPOWER2}. 
The base power of the IEEE 33-bus system is $100$ (MW) and the base voltage is $V_B= 230$ (kV).

The nonlinear dynamic models for Bus 18 and Bus 33 are summarized below.
Denote $\omega_{18}=\dot{\delta_{18}},\omega_{33}=\dot{\delta_{33}}$. The dynamic systems are
\bea M_{18} \dot \omega_{18} + g_{18}(\omega_{18})\ad = P^{18}_{in} - P^{18}_{out},\\
M_{33} \dot \omega_{33} + g_{33}(\omega_{33})\ad = P^{33}_{in} - P^{33}_{out},
\eea
where $P^i_{out}$ is the total transmitted power from Bus $i$ to its neighboring buses. Denote the line admittance $Y_{ij}=|Y_{ij}|\angle \gamma_{ij}$ and shunt admittance $Y_{i}=|Y_{i}|\angle \gamma_{i}$.

Since Bus 18 (and Bus 33) has only one neighboring Bus 17 (and Bus 32), we have
\begin{small}\bea
P^{18}_{out}= V^2_{18}|Y_{18}|\cos (\gamma_{18})+ V_{18}V_{17} |Y_{18,17}|\cos (\delta_{18}-\delta_{17} - \gamma_{18,17}),\eea
\end{small}
$\!\!$and similarly for Bus 33. The damping term is assumed to be linear with $b_{18}=0.22$, $b_{33}=0.12$. The normalized inertias are
$M_{18}=1.8$ and $M_{33}=0.9$.

Dynamic interactions of generators with the power grid are different from the traditional power flow analysis and introduce a new iteration scheme. During the transient time, $P^{18}_{in} \neq P^{18}_L+P^{18}_{out}$ which drives changes in $\delta_{18}$. The new $\delta_{18}$ then enters power flow analysis to result in new power flow status, including the new $P^{18}_{out}$; similarly for Bus 33.  As a result, during transient calculation of the power flow status, we designate the $66$ dependent variables in power flow calculation via MATPOWER as $Z=[P^d_{out}, Q^d_{out}, P_s, Q_s, V^{nd}, \delta^{nd}]$, where the superscript $d$ refers to the dynamic buses $18$ and $33$, $nd$ refers to the load buses $2-17$ and $19-32$, and the subscript $s$ refers to the slack bus $1$.

Under the generation powers $P^{18}_{in}=1.29$ pu and $P^{33}_{in}=0.89$ pu, the equilibrium point (the stationary operating condition) is calculated as $\bar \delta_{18}=-0.01$ (degree), $\bar \omega_{18}=0$, $\bar \delta_{33}=0.12$ (degree), $\bar \omega_{33}=0$.
The slack bus provides real power $3.94$ pu. The corresponding values of $Z$ at the equilibrium point are denoted by $\bar Z$.

Denote $x=(\delta_{18},\omega_{18},\delta_{33},\omega_{33})$, $u=[P^{18}_{in},P^{33}_{in}]$, the state equation is
$\dot x= f_0 (x, G(\delta_{18},\delta_{33}),u)=f(x, u).$
Then the Jacobian matrix at the equilibrium point $x=\bar x,Z=\bar Z$  is
\begin{small}
$$
A =  \frac{\partial f_0(x,Z,u)}{\partial x^{\mathrm{T}}} \Big|_{x=\bar x,Z=\bar Z}\! + \frac{\partial f_0(x,Z,u)}{\partial Z} \Big|_{x=\bar x,Z=\bar Z} \frac{\partial Z}{\partial  x^{\mathrm{T}}} \Big|_{x=\bar x,Z=\bar Z}.$$        
\end{small}

Based on the actual expressions of $f(x, u)$, the Jacobian matrix is given by
 \bea A = \begin{bmatrix} 0 & 1&0&0\\
-\frac{1}{M_{18}}
\frac{\partial P^{18}_{out}}{\partial\delta_{18}}&  -{b_{18}\over M_{18}} & -{1\over M_{18}}{\partial P^{18}_{out}\over \partial \delta_{33}}  & 0  \\
 0 &  0 & 0  &  1 \\
-{1\over M_{33}}{\partial P^{33}_{out}\over \partial \delta_{18}}&  0 & -{1\over M_{33}}{\partial P^{33}_{out}\over \partial \delta_{33}}  &  -{b_{33}\over M_{33}}
\end{bmatrix}.\eea

Utilizing the initial 33-bus power flow data sourced from \cite{Wu,Wong}\footnote{For the original data for the buses, links, generators, and loads of the 33-bus system,  please refer to the case33 file in MATPOWER, see \cite{MATPOWER1, MATPOWER2}.} and employing MATPOWER for power flow analysis along with the computation of partial derivatives, we  obtain \vspace{-3pt}
\bea
A=\begin{bmatrix}       0  &  1     &     0       &  0 \\
    -1.1280  & -0.1222  &  -0.0120      &   0\\
         0    &     0      &   0   & 1 \\
    -0.0344   &      0 &   -4.4785  & -0.1333
\end{bmatrix}.
\eea
 
\subsubsection{ Evaluation Scenarios and Input Design}

Let the line impedance of line $(i,j)$ be denoted by $X_{i,j}$.
Three cases are considered:
(1) Normal Operation: $\alpha=1: X_{1,2}=0.05753$ and $X_{26,27}=0.17732$, with probability $p_1=0.9$;
(2) Line Fault 1 (Increased Impedance of line (1,2)): $\alpha=2: X_{1,2}=0.5753$, with probability $p_2=0.06$;
(3) Line Fault 2 (Increased Impedance of line (26,27)): $\alpha=3: X_{26,27}=1.7732$, with probability $p_3=0.04$.
The corresponding $A(i)$ matrices are :
\begin{align}
&A(1)=\begin{bmatrix}
         0  &  1  &       0  &       0\\
   -1.1280  & -0.1222  & -0.0120  &       0\\
         0  &       0  &       0  &  1\\
   -0.0344  &       0  & -4.4785  & -0.1333
\end{bmatrix}, \nonumber\\ &A(2)=\begin{bmatrix}
         0&    1 &        0  &       0\\
   -1.1281&   -0.1222 &  -0.0127  &       0\\
         0&         0 &       0   & 1\\
   -0.0386&         0 &  -4.4877  & -0.1333
\end{bmatrix},\nonumber\\
&A(3)=\begin{bmatrix}
0  &   1 &         0   &      0\\
   -1.1277 &  -0.1222   &-0.0115     &    0\\
         0  &       0     &    0    &1\\
  -0.0299     & 0& -4.5179 &  -0.1333
\end{bmatrix}.\nonumber
\end{align}

We first calculate the eigenvalues of $A(i)$ and obtain their eigenvalues,
which are \begin{small}
    $\{-0.0611 + 1.0602i,
  -0.0611 - 1.0602i,
  -0.0667 + 2.1152i,
  -0.0667 - 2.1152i
\},$ $ \{   -0.0611 + 1.0603i,
  -0.0611 - 1.0603i,
  -0.0667 + 2.1174i,
  -0.0667 - 2.1174i\}$, $\{ -0.0611 + 1.0601i,
  -0.0611 - 1.0601i,
  -0.0667 + 2.1245i,
  -0.0667 - 2.1245i\}$
\end{small}.
Select  $u=0.1\sin t$ as the probing input,  which satisfies the conditions of Theorem \ref{thm3}.

\subsubsection{Results and Discussions} 
$$ $$
\vspace{-30pt}

\textbf{(i) Detection of Subsystems }

\begin{figure}[htbp] \vspace{-15pt}
	\setlength{\unitlength}{0.1 in}
	\begin{center}
		\includegraphics[height=5cm]{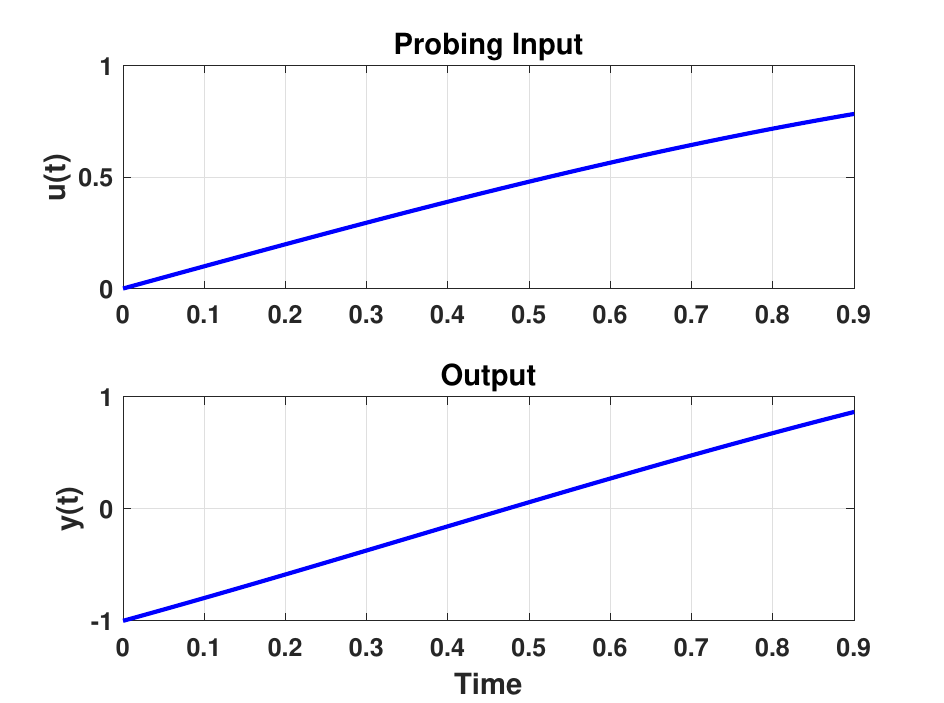} \vspace{-8pt}
	    \caption{The probing input $u(t)$ and the output $y(t)$ in $[0,\tau_0)$.}
    \label{fig33_0}
	\end{center}
\end{figure}

Take $\tau = 4.5$ and $\tau_0=0.9$. We show the detection of $\alpha_0$ for $t\in [0, \tau_0)$ as an example. Suppose the initial state is $[-1,2,1,2]$, and the true $\alpha_0=1.$ Now we estimate $\alpha_0$ by Algorithm 1. Under the probing input $u(t)=0.1\sin(t)$, Fig. \ref{fig33_0} shows the  curve of $u(t)$ and $y(t)$. According to Algorithm 1, we calculate the output prediction errors of the four subsystems, then we obtain $\varepsilon_1 = 7.5964\times 10^{-13},\ \varepsilon_2 =  8.1380\times 10^{-5},\ \varepsilon_3 =2.4539\times 10^{-4}.$ Therefore, $\wdh{\alpha}_0=1$, which detects the subsystem  accurately.

\textbf{(ii) Observer Design and Convergence}

We suppose that the sensor is on $\delta_{18}$, i.e., $C=[1,0,0,0].$
The pole placement design is used for designing observer feedback gains.
If we choose the desired closed-loop poles as $\lambda\!=\![-4, -3.2,  -4.8,-4.4]$, then the Matlab function $L_i=$ $ place(A^{\top}(i),C^{\top}, \lambda)$,  yields the suitable feedback gains and the closed-loop error dynamics with
$A_c^i=A(i)-L_i C$, $i=1,2,3$.

\begin{figure}[htbp] \vspace{-12pt}
	\setlength{\unitlength}{0.1 in}
	\begin{center}
		\includegraphics[height=5.5cm]{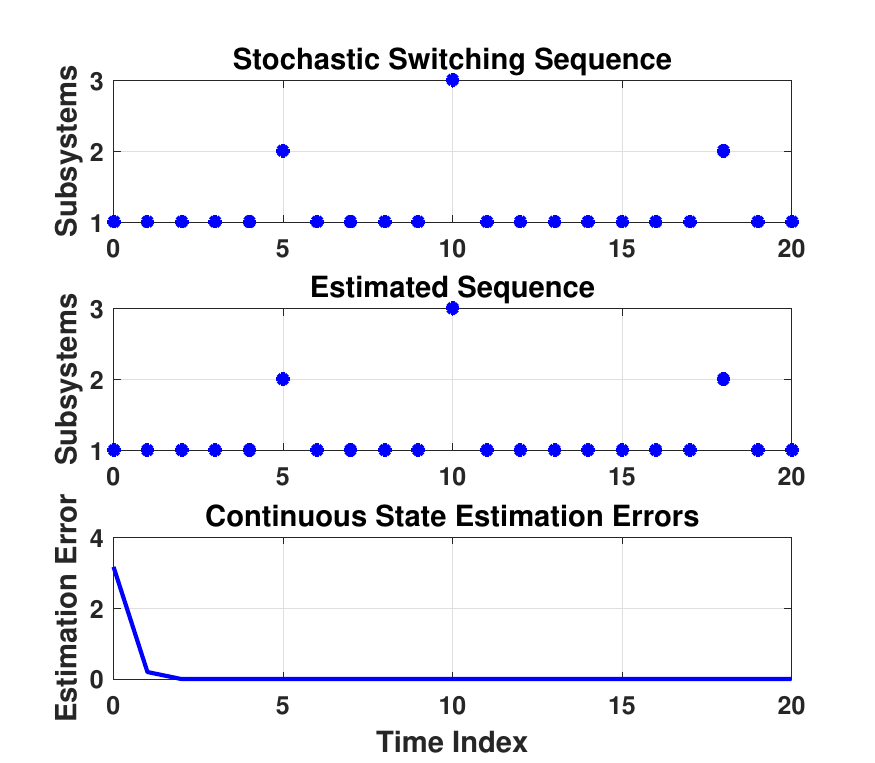} \vspace{-8pt}
	    \caption{The detection of $\alpha_k$ and the estimation error trajectory.}
    \label{fig33_1}
	\end{center}
\end{figure}

Take $\tau = 4.5$ and $\tau_0=0.9$.
The initial estimation error is selected to be $e(0)=[-1,2,1,2]^{\top}$ with the error norm $\sqrt{10}$.
Fig. \ref{fig33_1} shows that $\alpha_k$ can be accurately detected and the estimation error is convergent.
It should be pointed out that one sensor can identify different line faults.

\textbf{(iii) Packet Delivery}

Suppose that  $\delta_1$ and $\delta_2$ are independently measured.
Denote \begin{small}$C(1)=\begin{bmatrix}
   1 & 0 & 0 & 0\\  0 & 0 & 1 & 0  
\end{bmatrix}$\end{small} for normal operation, $C(2)=[1,0,0,0]$ for failure of Sensor 2 ($\delta_1$ measurement only), $C(3)=[0,0,1,0]$ for failure of Sensor 1 ($\delta_2$ measurement only), and $C(4)=[0,0,0,0]$ for failure on both sensors. Suppose that the packet delivery ratio for Sensor 1 is $\rho_1=0.95$ and for Sensor 2 is $\rho_2=0.97$. This data acquisition scheme can be modeled by an i.i.d. stochastic process $\alpha_k \in \clS=\{1,\ldots,4\}$ with $p_1=\rho_1 \rho_2=0.9215$, $p_2=\rho_1 (1-\rho_2)=0.0285$, $p_3=(1-\rho_1) \rho_2=0.0485$, $p_4=(1-\rho_1) (1-\rho_2)=0.0015$.

The pole placement design is used for designing observer feedback gains for $\alpha_k=1,2,3$. Since $C(4)=0$, the observer can only run open-loop.
For example, if we choose the desired closed-loop poles as $\lambda=[-1,-0.8,-1.2,-1.5]$, then the Matlab function $L_i= place(A',C'(i), \lambda)$,  yields the suitable feedback gains and the closed-loop error dynamics with
$A_c^i=A-L_i C(i)$, $i=1,2,3$.

\begin{figure}[htbp] \vspace{-12pt}
	\setlength{\unitlength}{0.1 in}
	\begin{center}
		\includegraphics[height=5.5cm]{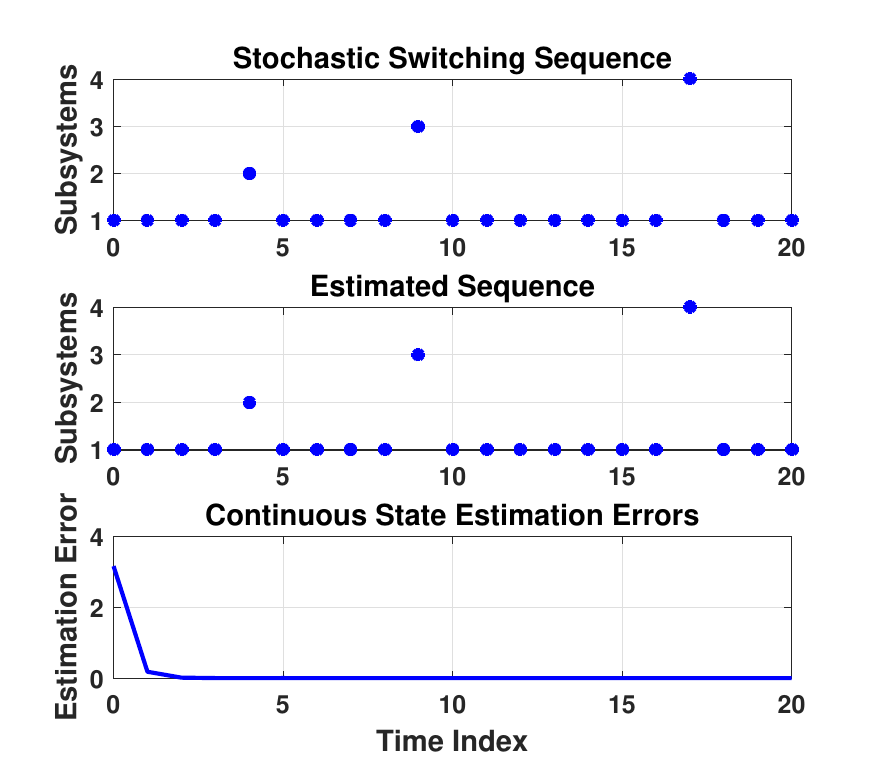} \vspace{-8pt}
	    \caption{The detection of $\alpha_k$ and the estimation error trajectory.}
    \label{fig33_3}
	\end{center}
\end{figure}

Take $\tau = 5$ and $\tau_0=1$.
The initial estimation error is selected to be $e(0)=[-1,2,1,2]^{\top}$ with the error norm $\sqrt{10}$.
Fig. \ref{fig33_3} shows that $\alpha_k$ can be accurately detected and the estimation error is convergent.

\section{Concluding Remarks}\label{Conc}

This paper introduced a new approach to contingency detection in MPS. The approach employs a stochastic hybrid system model in the state space form that captures both jumps from contingencies and internal dynamics of continuous states. Since system dynamics contain rich information on changes of system structures and parameters, the same output measurements for normal power system operation such as PMUs, frequencies, voltages, etc. can be used to detect contingencies of different types, causes, and locations. Model derivations, input design, detection algorithms, state estimation, and their stability and convergence properties were established. Practical bus systems were used to demonstrate the results.

\end{document}